\documentclass[journal]{IEEEtran}

\usepackage{epsfig,amsmath,amssymb,epsf,amsthm,scalefnt,multirow,subfig}
\usepackage{xcolor}
\usepackage{float}
\usepackage{cite}
\usepackage{psfrag}
\usepackage{graphics,amsmath,amssymb,graphicx,amsthm,scalefnt,multirow,dsfont,subfig}

\usepackage{epsfig,amsmath,amssymb,epsf,amsthm,scalefnt,multirow}
\usepackage{xcolor}
\usepackage{float}
\usepackage{cite}
\usepackage{psfrag}
\usepackage{cleveref}
\usepackage{mathrsfs}
\usepackage{mathtools}
\usepackage{algpseudocode}
\usepackage{algorithm}
\usepackage{enumerate}
\usepackage{stfloats}

\newtheorem{theorem}{Theorem}
\newtheorem{lemma}{Lemma}
\newtheorem*{lemma*}{Lemma}

\newtheorem{proposition}{Proposition}

  \def\cC{{\mathcal{C}}}

\def\cM{{\mathcal{M}}} \def\cN{{\mathcal{N}}} \def\cO{{\mathcal{O}}} 
 \def\cR{{\mathcal{R}}}  \def\cT{{\mathcal{T}}}

\def\diag{\mathop{\mathrm{diag}}}

\def\trace{\mathop{\mathrm{tr}}}

\def\Re{\mathop{\mathrm{Re}}}

\def\bSigma{{\pmb{\Sigma}}}

\def\bTheta{{\pmb{\Theta}}} \def\btheta{{\pmb{\theta}}}
 
 \def\bmu{{\pmb{\mu}}}
\def\b0{{\pmb{0}}}\def\bLambda{{\pmb{\Lambda}}} 

\def\ba{{\mathbf{a}}}  \def\bc{{\mathbf{c}}} \def\bd{{\mathbf{d}}}
  \def\bg{{\mathbf{g}}} \def\bh{{\mathbf{h}}}
   
 \def\bn{{\mathbf{n}}}  
  \def\bs{{\mathbf{s}}} 
   \def\bx{{\mathbf{x}}}
\def\by{{\mathbf{y}}}   

\def\bA{{\mathbf{A}}} \def\bB{{\mathbf{B}}} \def\bC{{\mathbf{C}}} \def\bD{{\mathbf{D}}}
\def\bE{{\mathbf{E}}} \def\bF{{\mathbf{F}}} \def\bG{{\mathbf{G}}} \def\bH{{\mathbf{H}}}
\def\bI{{\mathbf{I}}}  \def\bK{{\mathbf{K}}} 
   \def\bP{{\mathbf{P}}}
   
\def\bU{{\mathbf{U}}} \def\bV{{\mathbf{V}}}  \def\bX{{\mathbf{X}}}
\def\bY{{\mathbf{Y}}}

\begin{document}	
\bstctlcite{IEEEexample:BSTcontrol}
\title{Joint Downlink and Uplink Optimization for RIS-Aided FDD MIMO Communication Systems}

\author{\IEEEauthorblockN{Gyoseung Lee, Hyeongtaek Lee, Donghwan Kim, Jaehoon Chung, A. Lee. Swindlehurst, and Junil Choi}
\thanks{This work was supported in part by LG Electronics Inc.; in part by Basic Science Research Program through the National Research Foundation of Korea (NRF) funded by the Ministry of Education (RS-2023-00271715); in part by the Ministry of Science and ICT (MSIT), South Korea, under the Information Technology Research Center (ITRC) Support Program supervised by the Institute of Information and Communications Technology Planning and Evaluation (IITP) under Grant IITP-2020-0-01787; and in part by the Korea Institute for Advancement of Technology (KIAT) grant funded by the Ministry of Trade, Industry and Energy (MOTIE) (P0022557). The work of A.~Swindlehurst was supported by the U.S. National Science Foundation under grants ECCS-2030029 and CNS-2107182.}
\thanks{Gyoseung Lee, Hyeongtaek Lee, and Junil Choi are with the School of Electrical Engineering, Korea Advanced Institute of Science and Technology (e-mail: \{iee4432; htlee8459; junil\}@kaist.ac.kr).}
\thanks{Donghwan Kim and Jaehoon Chung are with C\&M Standard Lab, ICT Technology Center, LG Electronics Inc. (e-mail: \{donghwan88.kim; jaehoon.chung\}@lge.com).}
\thanks{A. L. Swindlehurst is with the Center for Pervasive Communications and Computing, Henry Samueli School of Engineering, University of California, Irvine, CA 92697, USA (e-mail: swindle@uci.edu).}}

\maketitle

\begin{abstract}
This paper investigates reconfigurable intelligent surface (RIS)-aided frequency division duplexing (FDD) communication systems.
Since the downlink and uplink signals are simultaneously transmitted in FDD, the phase shifts at the RIS should be designed to support both transmissions.
Considering a single-user multiple-input multiple-output system, we formulate a weighted sum-rate maximization problem to jointly maximize the downlink and uplink system performance.
To tackle the non-convex optimization problem, we adopt an alternating optimization (AO) algorithm, in which two phase shift optimization techniques are developed to handle the unit-modulus constraints induced by the reflection coefficients at the RIS.
The first technique exploits the manifold optimization-based algorithm, while the second uses a lower-complexity AO approach.
Numerical results verify that the proposed techniques rapidly converge to local optima and significantly improve the overall system performance compared to existing benchmark~schemes.
\end{abstract}


\section{Introduction}\label{sec1}
Reconfigurable intelligent surfaces (RISs), which consist of a planar metamaterial structure, have recently emerged as a promising candidate for future wireless communication systems \cite{Wu:2020, Renzo:2019, Basar:2019, Renzo:2022, Huang:2018, Huang:2020}. 
With its ability to dynamically control the amplitude and/or phase of incoming signals, an RIS can modify the signal propagation and lead to enhanced spectral efficiency and reduced power consumption \cite{Wu:2020, Renzo:2019, Basar:2019, Renzo:2022}.
For instance, when the direct link channel between the base station (BS) and user equipment (UE) is obstructed, the RIS can establish a virtual BS-RIS-UE link, thereby improving the coverage of wireless communication systems \cite{Wu:2020, Basar:2019}.

When it comes to designing RIS-aided communication systems, existing works have mainly focused on time division duplexing (TDD) to leverage the channel reciprocity between the downlink and uplink channels \cite{Pan:2021}.
In \cite{Wu:2019}, transmit power minimization strategies were developed for the joint design of active and passive beamforming. 
In \cite{Guo:2020}, a weighted sum-rate maximization problem was formulated for multi-user multiple-input single-output (MU-MISO) systems.
A low-complexity algorithm with a two-timescale transmission protocol was developed in \cite{Reviewer_1_1} to maximize the achievable weighted sum-rate for RIS-aided cell-free systems.
In \cite{Reviewer_1_2}, ergodic sum capacity maximization strategies for the downlink and uplink were developed.
Furthermore, several works have considered single-user multiple-input multiple-output (SU-MIMO) TDD systems \cite{Zhang:2020, Wang:2021, Hong:2022, ICL_TWC, Bahingayi:2022, Li:2023}.
In \cite{Zhang:2020}, downlink capacity maximization strategies were investigated, and the reflection coefficients at the RIS were optimized in an alternating manner.
Lower-complexity optimization techniques were developed in \cite{Wang:2021} by approximating the singular values of millimeter-wave (mmWave) channels in terms of the reflection coefficients.
In \cite{Hong:2022}, hybrid beamformers and reflection coefficients were designed by exploiting the structure of mmWave systems in the asymptotic regime where the number of antennas at the BS and UE and the number of RIS elements go to infinity.
A reflection coefficient design using only linear operations was developed in \cite{ICL_TWC}, and array selection algorithms were developed in \cite{Bahingayi:2022} to maximize the capacity of each RIS-related link.
In \cite{Li:2023}, strategies that maximize the ergodic achievable rate were investigated.

Despite the advantages of TDD in designing RIS-aided communication systems, sub-6 GHz bands in future wireless communication systems will still remain significant due to their broad coverage and reliability \cite{Viswanathan:2020}. This suggests that various future applications may still rely on frequency division duplexing (FDD), and RISs would inevitably be deployed in FDD systems as well.
Consequently, there is a need to study RIS-aided FDD systems, and some recent RIS-related studies have attempted to exploit the structure of FDD \cite{JSDM:FDD, Wang:2022, Guo:2021, Zhou:2023, Abouamer:2022}. 
Focusing on the downlink design, joint spatial division and multiplexing approaches based on statistical channel state information (CSI) were investigated in \cite{JSDM:FDD}, and a Bayesian optimization-based beamforming strategy without CSI feedback was proposed in~\cite{Wang:2022}. While existing phase shift optimization algorithms that focus on TDD-based SU-MIMO systems can be directly applied to optimize the reflection coefficients for either the downlink or uplink, we will see that designing for only one direction will generally lead to highly suboptimal performance for the other.

One important characteristic of RIS-aided FDD systems is that the transmissions of downlink and uplink signals occur simultaneously in different frequency bands.
In conventional FDD systems without RISs, it is difficult to satisfy the different demands on the downlink and uplink rates due to the use of fixed system bandwidths. However, deploying RISs enable one to strike a balance the between the downlink and uplink transmissions to satisfy such demands, implying that beyond the individual downlink or uplink design, the reflection coefficients at the RIS should be optimized to enhance the overall downlink and uplink system performance. A few recent works have addressed this challenge \cite{Guo:2021, Zhou:2023, Abouamer:2022}.
In \cite{Guo:2021}, a joint optimization framework was developed for single-user multiple-input single-output (SU-MISO) systems.
An equivalent circuit model was introduced in \cite{Zhou:2023} to address practical RIS design in SU-MISO systems.
In \cite{Abouamer:2022}, joint resource allocation strategies for MU-MISO systems were proposed. 
However, there is no prior work that considers the case of a multi-antenna UE joint downlink and uplink optimization in FDD systems.

In this paper, we focus on the FDD SU-MIMO scenario and propose a framework to maximize the weighted sum-rate for the downlink and uplink using a weight to control the relative priority of the downlink and uplink transmissions.
By properly setting the weight coefficient, various system requirements on the relative downlink and uplink rates can be achieved.
To tackle the resulting non-convex optimization problem, we first decouple the design of the RIS reflection coefficients from that of the transmit precoders at the BS and UE. Subsequently, we concentrate on the optimization problem for the RIS reflection coefficients and propose two techniques to handle the unit-modulus constraints, which are the main challenge for obtaining a practical solution for RIS-aided systems.
The first proposed technique employs a manifold optimization-based algorithm that exploits the fact that the reflection coefficients at the RIS lie on a complex circle manifold.
In the second, we propose a lower-complexity alternating optimization (AO) algorithm in which the sub-problem corresponding to one reflection coefficient is formulated while keeping all other variables fixed,
and the reflection coefficients are optimized by iteratively solving these sub-problems. Closed-form solutions are derived for each sub-problem, making the proposed algorithm computationally efficient compared to the manifold optimization-based~approach.

Our numerical results verify the convergence of the proposed algorithms, and we demonstrate that the proposed joint downlink and uplink designs significantly improve the overall system performance compared to existing benchmarks.
The downlink and uplink rate regions with respect to the weight coefficients are also investigated to verify the effectiveness of the proposed joint optimization framework.

The rest of this paper is organized as follows.
In Section \ref{sec2}, the system model for the assumed RIS-aided FDD SU-MIMO system is presented.
In Section \ref{sec3}, the problem formulation and algorithm design are investigated, and the two phase shift optimization techniques are proposed in Section \ref{sec4}. Numerical results for the proposed algorithms are provided in Section \ref{sec5}, and we conclude the paper in Section \ref{conclusion}.

$Notation$: Lower and upper boldface letters represent column vectors and matrices. The element-wise conjugate, transpose, and conjugate transpose of a matrix $\bA$ are denoted by $\bA^*$, $\bA^{\mathrm{T}}$, and $\bA^{\mathrm{H}}$, respectively.
For a square matrix $\bA$, $\det(\bA)$, $\trace(\bA)$, and $\bA^{-1}$ are respectively the determinant, trace, and inverse of $\bA$.
The quantities $\bA(i,:)$ and $[\bA]_{i,j}$ denote the $i$-th row and the $(i,j)$-th entry of the matrix $\bA$, respectively. The notation
$\mathrm{diag}(\ba)$ represents a diagonal matrix whose diagonal elements correspond to the entries of the vector $\ba$.
The $\ell_2$-norm of a vector $\ba$ and the Frobenius-norm of a matrix $\bA$ are respectively denoted by $\Vert \ba \Vert_2$ and $\Vert \bA \Vert_{\mathrm{F}}$.
A circularly symmetric complex Gaussian distribution with mean vector $\bmu$ and covariance matrix $\bK$ is represented using $\cC\cN(\bmu,\bK)$. The quantities
$\b0_{m}$, $\b0_{m,n}$, and $\bI_m$ represent the $m \times 1$ all-zero vector, the $m \times n$ all-zero matrix, and the $m \times m$ identity matrix, respectively. 
The expressions $\vert a \vert$, $a^*$, $\arg(a)$, and $\Re(a)$ represent the magnitude, conjugate, angle, and real part of a complex number $a$, respectively.
The Kronecker product is defined as $\otimes$, and $\circ$ denotes the Hadamard product. Standard ``big-O'' notation is indicated by~$\cO(\cdot)$.

\section{System Model}\label{sec2}
\begin{figure}
	\centering
	\includegraphics[width=1.02\columnwidth]{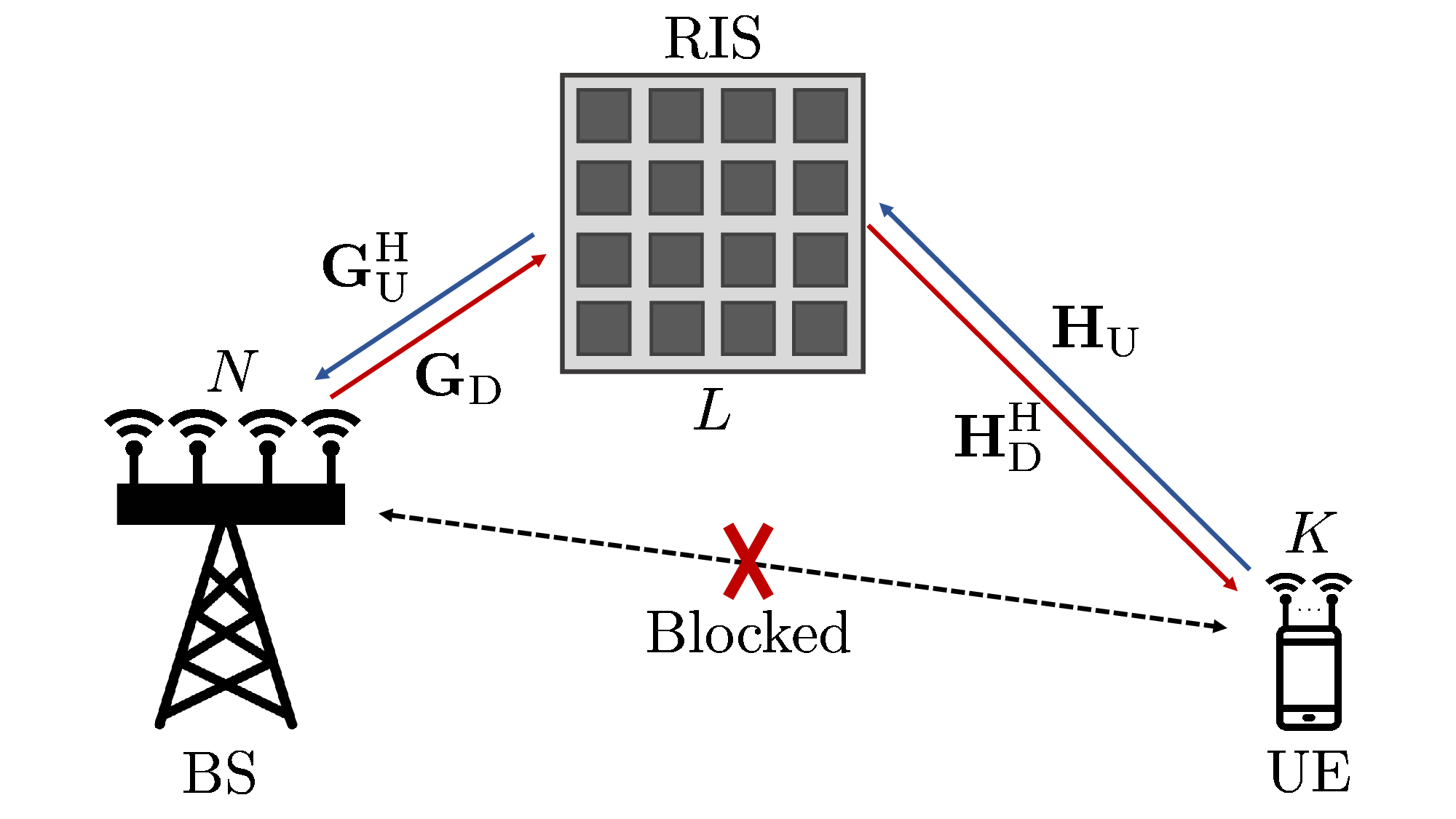}
	\caption{An example of an RIS-aided FDD SU-MIMO communication system with $N$ BS antennas, $K$ UE antennas, and $L$ RIS elements.}	\label{system_model}
\end{figure}

We consider the RIS-aided FDD SU-MIMO scenario depicted in Fig. \ref{system_model}, where the BS deploys $N$ antennas and serves the UE with $K$ antennas.
The RIS consists of $L$ purely passive elements. The RIS is connected to the BS via a controller that allows the BS to control the RIS elements to achieve the desired signal reflection.

As in \cite{Guo:2021,Wang:2021}, the direct link channels between the BS and UE are assumed to be completely blocked by an obstacle. In the downlink transmission, the BS transmits $N_{\mathrm{s}}^{\mathrm{D}}$ data streams to the UE where $N_{\mathrm{s}}^{\mathrm{D}}\leq \min(N,K)$.
Let the signal vector transmitted from the BS be $\bs_{\mathrm{D}} \in \mathbb{C}^{N_{\mathrm{s}}^{\mathrm{D}} \times 1}$, which satisfies $\mathbb{E}\left[ \bs_{\mathrm{D}} \bs_{\mathrm{D}}^{\mathrm{H}}\right] = \bI_{N_{\mathrm{s}}^{\mathrm{D}}}$. 
The received signal at the UE is then given~by 
\begin{equation} \label{y_D}
	\by_{\mathrm{D}} = \bH_{\mathrm{D}}^{\mathrm{H}}\bTheta\bG_{\mathrm{D}} \bF_{\mathrm{D}} \bs_{\mathrm{D}} + \bn_{\mathrm{D}},
\end{equation}
where $\bF_{\mathrm{D}} \in \mathbb{C}^{N \times N_{\mathrm{s}}^{\mathrm{D}}}$ represents the precoding matrix employed at the BS satisfying $\Vert \bF_{\mathrm{D}} \Vert_{\mathrm{F}}^2 \leq P_{\mathrm{D,max}}$ with maximum downlink transmit power $P_{\mathrm{D,max}}$, $\bH_{\mathrm{D}}^{\mathrm{H}} \in \mathbb{C}^{K \times L}$ is the channel from the RIS to the UE, $\bG_{\mathrm{D}} \in \mathbb{C}^{L \times N}$ is the channel from the BS to the RIS, and $\bn_{\mathrm{D}} \sim \cC\cN(\b0_{K}, \sigma_{\mathrm{D}}^2 \bI_{K})$ is an additive white Gaussian noise (AWGN) vector at the UE with noise variance $\sigma_{\mathrm{D}}^2$. The $L \times L$ reflection coefficient matrix of the RIS is defined as $\bTheta= \mathrm{diag}\left([\theta_1,\cdots,\theta_L]^{\mathrm{T}}\right)$ with $\vert \theta_{\ell} \vert=1$, $\ell=1,\cdots,L$. 

Similarly, in the uplink transmission, the UE transmits $N_{\mathrm{s}}^{\mathrm{U}}$ data streams to the BS where $N_{\mathrm{s}}^{\mathrm{U}} \leq \min(N,K)$. Let $\bs_{\mathrm{U}} \in \mathbb{C}^{N_{\mathrm{s}}^{\mathrm{U}} \times 1}$ be the signal vector sent from the UE satisfying $\mathbb{E}\left[ \bs_{\mathrm{U}} \bs_{\mathrm{U}}^{\mathrm{H}}\right] =  \bI_{N_{\mathrm{s}}^{\mathrm{U}}}$.
Note that, due to its design, the reflection response of the RIS varies with frequency \cite{Reviewer_3_1, Reviewer_2_1, Wenhao:2020, Qunqiang:2023}. Nevertheless, if the RIS elements are properly designed and the gap between the two frequencies is not too large, it is possible to achieve a constant phase offset in the RIS response at the two frequencies \cite{Qunqiang:2023}.
Taking the frequency spectrum of LTE band 1 as an example, the frequency ranges for the downlink and uplink are respectively 2.11-2.17 GHz and 1.92-1.98 GHz, which are not too far apart \cite{LTE_1_band, JSDM:FDD}.
Hence, we assume the reflection coefficient matrix in the uplink channel is $\widetilde{\bTheta}=e^{j\theta_{\mathrm{d}}} \bTheta$, where $\theta_{\mathrm{d}} \in [0,2\pi)$ denotes the bulk phase difference in the reflection response between the downlink and uplink. Then, the received signal at the BS is represented~by
\begin{align}
	\by_{\mathrm{U}} &= \bG_{\mathrm{U}}^{\mathrm{H}}\widetilde{\bTheta}\bH_{\mathrm{U}} \bF_{\mathrm{U}} \bs_{\mathrm{U}} + \bn_{\mathrm{U}} \nonumber 
    \\ &= e^{j\theta_{\mathrm{d}}}\bG_{\mathrm{U}}^{\mathrm{H}}\bTheta\bH_{\mathrm{U}} \bF_{\mathrm{U}} \bs_{\mathrm{U}} + \bn_{\mathrm{U}},
\end{align}
where $\bF_{\mathrm{U}} \in \mathbb{C}^{K \times N_{\mathrm{s}}^{\mathrm{U}}}$ denotes the precoding matrix employed at the UE satisfying $\Vert \bF_{\mathrm{U}} \Vert_{\mathrm{F}}^2 \leq P_{\mathrm{U,max}}$ with maximum uplink transmit power $P_{\mathrm{U,max}}$, $\bG_{\mathrm{U}}^{\mathrm{H}} \in \mathbb{C}^{N \times L}$ is the channel from the RIS to the BS, $\bH_{\mathrm{U}} \in \mathbb{C}^{L \times K}$ is the channel from the UE to the RIS, and $\bn_{\mathrm{U}} \sim \cC\cN(\b0_{N}, \sigma_{\mathrm{U}}^2 \bI_{N})$ is an AWGN vector at the BS with noise variance $\sigma_{\mathrm{U}}^2$. Since our proposed techniques do not rely on any specific channel model, we do not specify the BS-RIS and RIS-UE channels in this section. Specific choices for these channels based on the popular geometric channel model will be given to describe the simulation scenarios in Section~\ref{sec5_1}.

By defining the effective downlink and uplink channels as $\bH_{\mathrm{eff,D}}=\bH_{\mathrm{D}}^{\mathrm{H}}\bTheta\bG_{\mathrm{D}}$ and $\widetilde{\bH}_{\mathrm{eff,U}}=e^{j\theta_{\mathrm{d}}}\bG_{\mathrm{U}}^{\mathrm{H}}\bTheta\bH_{\mathrm{U}}=e^{j\theta_{\mathrm{d}}}\bH_{\mathrm{eff,U}}$, the downlink achievable rate $R_{\mathrm{D}}$ and uplink achievable rate $R_{\mathrm{U}}$ are given by
\begin{align}
	R_{\mathrm{D}} &= \log_2 \det \left(\bI_{K}+ \frac{1}{\sigma_{\mathrm{D}}^2}  \bH_{\mathrm{eff,D}} \bF_{\mathrm{D}} \bF_{\mathrm{D}}^{\mathrm{H}} \bH_{\mathrm{eff,D}}^{\mathrm{H}}  \right), \nonumber \\
	R_{\mathrm{U}} &= \log_2 \det \left(\bI_{N}+ \frac{1}{\sigma_{\mathrm{U}}^2}  \widetilde{\bH}_{\mathrm{eff,U}} \bF_{\mathrm{U}} \bF_{\mathrm{U}}^{\mathrm{H}} \widetilde{\bH}_{\mathrm{eff,U}}^{\mathrm{H}} \right) \nonumber
    \\ &= \log_2 \det \left(\bI_{N}+ \frac{1}{\sigma_{\mathrm{U}}^2}  \bH_{\mathrm{eff,U}} \bF_{\mathrm{U}} \bF_{\mathrm{U}}^{\mathrm{H}} \bH_{\mathrm{eff,U}}^{\mathrm{H}} \right). \label{Achievable_rate}
\end{align}

\section{Problem Formulation and Algorithm Design} \label{sec3}
\subsection{Problem formulation}
	In FDD, due to the simultaneously transmitted downlink and uplink signals, the precoders at the BS and UE and reflection coefficients at the RIS should be designed to jointly maximize the downlink and uplink rates. In (\ref{Achievable_rate}) we observe that the same reflection coefficient matrix $\bTheta$ appears in both $R_{\mathrm{D}}$ and $R_{\mathrm{U}}$ irrespective of the value of $\theta_{\mathrm{d}}$.
	Therefore, we formulate the following weighted sum-rate maximization problem for the downlink and uplink as \cite{Guo:2021, Abouamer:2022}
\begin{align}
	(\mathrm{P1}): \max_{\bF_{\mathrm{D}},\bF_{\mathrm{U}},\bTheta} & \enspace R_{\mathrm{WSR}}=\eta R_{\mathrm{D}}+(1-\eta) R_{\mathrm{U}} \label{objective_P1} \\ \mbox{s.t. } \quad & \enspace \Vert \bF_{\mathrm{D}} \Vert_{\mathrm{F}}^2 \leq P_{\mathrm{D,max}}, \\ \quad \quad \enspace  & \enspace \Vert \bF_{\mathrm{U}}\Vert_{\mathrm{F}}^2  \leq P_{\mathrm{U,max}}, \\ \quad \quad \enspace & \enspace \bTheta= \mathrm{diag}\left([\theta_1,\cdots,\theta_L]^{\mathrm{T}}\right), \\\quad \quad  & \enspace \vert \theta_{\ell} \vert =1, \enspace \ell=1,\cdots,L, \label{P1_phase_constraint} 
\end{align}
where $\eta \in [0,1]$ is the weight coefficient that controls the relative priority between the downlink and uplink.

The objective function (\ref{objective_P1}) is the weighted sum of non-concave functions $R_{\mathrm{D}}$ and $R_{\mathrm{U}}$ with respect to the reflection coefficient matrix $\bTheta$, and the unit-modulus constraints in (\ref{P1_phase_constraint}) are non-convex. Furthermore, the precoding matrices $\bF_{\mathrm{D}}$ and $\bF_{\mathrm{U}}$ are coupled with $\bTheta$, making the optimization problem (P1) difficult to solve. To address these issues, we adopt the AO algorithm. For a fixed $\bTheta$, problem (P1) can be decomposed into downlink and uplink sub-problems, and the optimal transmit precoders can be obtained using eigenmode transmissions \cite{Fundamental_Tse}. For fixed $\bF_{\mathrm{D}}$ and $\bF_{\mathrm{U}}$, (P1) must be solved with respect to $\bTheta$ alone, and we propose two optimization techniques to tackle the unit-modulus constraints.

\subsection{Optimization of $\bF_{\mathrm{D}}$ and $\bF_{\mathrm{U}}$ with given $\bTheta$}
When $\bTheta$ is fixed, (P1) can be decoupled into separate downlink and uplink sub-problems.
The downlink sub-problem for (P1) is given by
\begin{align}
	\max_{\bF_{\mathrm{D}}} \enspace &R_{\mathrm{D}} \nonumber
	\\  \mbox{s.t.} \enspace \enspace  &\Vert \bF_{\mathrm{D}} \Vert_{\mathrm{F}}^2 \leq P_{\mathrm{D,max}}.
\end{align}
In this problem, the optimal $\bF_{\mathrm{D}}$ can be obtained by the singular value decomposition (SVD) of the effective downlink channel and the water-filling power allocation. Denoting the truncated SVD of the effective downlink channel as  $\bH_{\mathrm{eff,D}}=\tilde{\bU}_{\mathrm{D}}\tilde{\bSigma}_{\mathrm{D}}\tilde{\bV}_{\mathrm{D}}^{\mathrm{H}}$ with $\tilde{\bV}_{\mathrm{D}} \in \mathbb{C}^{N \times N_{\mathrm{s}}^{\mathrm{D}}}$, the optimal $\bF_{\mathrm{D}}$ is given by 
\begin{equation} \label{F_D_opt}
	\bF_{\mathrm{D}}^{\star}=\tilde{\bV}_{\mathrm{D}} \bP_{\mathrm{D}}^{\frac{1}{2}},
\end{equation}
where $\bP_{\mathrm{D}} = \diag\left( \left[p^{\star}_{\mathrm{D},1}, \cdots, p^{\star}_{\mathrm{D},N_{\mathrm{s}}^{\mathrm{D}}} \right]^{\mathrm{T}} \right)$ denotes the downlink power allocation matrix with $p_{\mathrm{D},i}^{\star}$ representing the optimal amount of power allocated to the $i$-th data stream obtained by the water-filling power allocation. Specifically, $p_{\mathrm{D},i}^{\star}$ is given by $p_{\mathrm{D},i}^{\star}=\max(1/p_{\mathrm{D},0}-\sigma_{\mathrm{D}}^2/[\tilde{\bSigma}_{\mathrm{D}}]_{i,i}, 0)$ for $i=1,\cdots, N_{\mathrm{s}}^{\mathrm{D}}$, where $p_{\mathrm{D},0}$ satisfies $\sum_{i=1}^{N_{\mathrm{s}}^{\mathrm{D}}} p_{\mathrm{D},i}^{\star} = P_{\mathrm{D,max}}$.

Similarly, the uplink sub-problem for (P1) is given by
\begin{align}
	\max_{\bF_{\mathrm{U}}} \enspace &R_{\mathrm{U}} \nonumber 
	\\  \mbox{s.t.} \enspace \enspace  &\Vert \bF_{\mathrm{U}} \Vert_{\mathrm{F}}^2 \leq P_{\mathrm{U,max}}.
\end{align}
The optimal $\bF_{\mathrm{U}}$ can be computed as in (\ref{F_D_opt}), i.e.,
\begin{equation} \label{F_U_opt}
	\bF_{\mathrm{U}}^{\star}=\tilde{\bV}_{\mathrm{U}} \bP_{\mathrm{U}}^{\frac{1}{2}},
\end{equation}
where $\tilde{\bV}_{\mathrm{U}} \in \mathbb{C}^{K \times N_{\mathrm{s}}^{\mathrm{U}}}$ is the right singular matrix of the truncated SVD $\bH_{\mathrm{eff,U}}=\tilde{\bU}_{\mathrm{U}}\tilde{\bSigma}_{\mathrm{U}}\tilde{\bV}_{\mathrm{U}}^{\mathrm{H}}$, and $\bP_{\mathrm{U}}=\diag\left( \left[p^{\star}_{\mathrm{U},1}, \cdots, p^{\star}_{\mathrm{U},N_{\mathrm{s}}^{\mathrm{U}}}\right]^{\mathrm{T}} \right)$ is the uplink power allocation matrix obtained by the water-filling power allocation. The uplink power allocated for the $i$-th data stream is given by $p_{\mathrm{U},i}^{\star}=\max(1/p_{\mathrm{U},0}-\sigma_{\mathrm{U}}^2/[\tilde{\bSigma}_{\mathrm{U}}]_{i,i}, 0)$ for $i=1,\cdots, N_{\mathrm{s}}^{\mathrm{U}}$, where $p_{\mathrm{U},0}$ satisfies $\sum_{i=1}^{N_{\mathrm{s}}^{\mathrm{U}}} p_{\mathrm{U},i}^{\star} = P_{\mathrm{U,max}}$.

\subsection{Optimization of $\bTheta$ with given $\bF_{\mathrm{D}}$ and $\bF_{\mathrm{U}}$}

For fixed $\bF_{\mathrm{D}}$ and $\bF_{\mathrm{U}}$, (P1) must be optimized with respect to the reflection coefficients at the RIS:
\begin{align}
	\mbox{(P2)}: \enspace \max_{\bTheta} \enspace &R_{\mathrm{WSR}} \label{Phase_shift_objective} \\ \mbox{s.t. } \enspace &\bTheta= \mathrm{diag}\left([\theta_1,\cdots,\theta_L]^{\mathrm{T}}\right), \\ \quad \quad  &\vert \theta_{\ell} \vert =1, \enspace \ell=1,\cdots,L. \label{phase_constraint}
\end{align}
The non-convex constraints (\ref{phase_constraint}) still make it challenging to find the optimal solution for this problem. 
We tackle this issue in the following section.

\section{Proposed Phase Shift Optimization Techniques}\label{sec4}

In this section, we propose two techniques to optimize the reflection coefficients at the RIS in (P2) to simultaneously improve the downlink and uplink system performance.
In the first technique, we adopt a manifold optimization-based algorithm by leveraging the fact that the reflection coefficients at the RIS lie on the complex circle manifold.
For the second, we develop a lower-complexity AO technique in which an effective closed-form solution for each reflection coefficient is derived. The relative computational complexity of the two methods will be compared.

\subsection{Manifold optimization}\label{sec4_1}

As discussed in the previous section, the main obstacle to solving (P2) is the presence of unit-modulus constraints in (\ref{phase_constraint}), making the optimization problem highly non-convex. Fortunately, these unit-modulus constraints form the complex circle manifold $\cM_{cc}^L = \{ \btheta \in \mathbb{C}^{L} : \vert \theta_1 \vert = \cdots = \vert \theta_L \vert=1 \}$ \cite{Book:manifold},  which is a Riemannian manifold. The key advantage of dealing with the Riemannian manifold is that optimization methods applicable in the Euclidean space, such as gradient descent, can also be employed on Riemannian manifolds. Therefore, we adopt the Riemannian conjugate gradient (RCG) algorithm to obtain a stationary point for (P2). The RCG algorithm is the generalization of the conjugate gradient method to the Riemannian manifold space, and it can efficiently tackle optimization problems with non-convex constraints such as the unit-modulus constraints \cite{Guo:2021,Wang:2021,Li:2023}. To implement the RCG-based algorithm, the following three steps must be implemented.

\subsubsection{Compute Riemannian gradient}

The Riemannian gradient corresponds to the direction of maximum ascent of the objective function at the point $\btheta=\left[\theta_1,\cdots,\theta_L\right]^{\mathrm{T}} \in \cM_{cc}^{L}$, while being restricted within its tangent space. It can be computed by the orthogonal projection of the Euclidean gradient onto the tangent space. On the complex circle manifold, the Riemannian gradient of objective function (\ref{Phase_shift_objective}) is expressed~as
\begin{align} \label{Riemannian_grad}
	&\operatorname{grad}_{\btheta} R_{\mathrm{WSR}}\nonumber \\& = \nabla_{\btheta} R_{\mathrm{WSR}} - \Re{\left(\nabla_{\btheta} R_{\mathrm{WSR}} \circ \btheta^* \right)} \circ \btheta,
\end{align}
where $\nabla_{\btheta} R_{\mathrm{WSR}}$ represents the Euclidean gradient of $R_{\mathrm{WSR}}$ with respect to $\btheta$. To compute the Riemannian gradient (\ref{Riemannian_grad}), $\nabla_{\btheta} R_{\mathrm{WSR}}=\eta \nabla_{\btheta} R_{\mathrm{D}} + (1-\eta)\nabla_{\btheta} R_{\mathrm{U}}$ must be found. Note that once $\nabla_{\btheta} R_{\mathrm{D}}=\left[ \frac{\partial R_{\mathrm{D}}}{\partial \theta_1} , \cdots, \frac{\partial R_{\mathrm{D}}}{\partial \theta_L} \right]^{\mathrm{T}}$ is obtained, the computation of $\nabla_{\btheta} R_{\mathrm{U}}$ is straightforward. By applying the chain rule in \cite{Palomar:2006}, the partial derivative of $R_{\mathrm{D}}$ with respect to the $\ell$-th reflection coefficient $\theta_{\ell}$ can be represented by
\begin{align} \label{Partial_R_D_theta_ell}
	 \frac{\partial R_{\mathrm{D}}}{\partial \theta_{\ell}} =&\trace \left( \nabla_{\bH_{\mathrm{eff,D}}} R_{\mathrm{D}} \cdot \frac{\partial \bH_{\mathrm{eff,D}}^{\mathrm{H}}}{\partial \theta_{\ell}^*} \right) \nonumber \\&+ \trace \left( (\nabla_{\bH_{\mathrm{eff,D}}} R_{\mathrm{D}})^{\mathrm{H}} \cdot \frac{\partial \bH_{\mathrm{eff,D}}}{\partial \theta_{\ell}^*}  \right).
\end{align}
In the complex differentials, $\theta_{\ell}$ and $\theta_{\ell}^*$ can be treated as independent variables \cite{Hjorungenes:2007}, from which it is obvious that $\frac{\partial \bH_{\mathrm{eff,D}}}{\partial \theta_{\ell}^*}=0$, and only computation of the first term in (\ref{Partial_R_D_theta_ell}) is required. We first derive the Euclidean gradient of $R_{\mathrm{D}}$ with respect to $\bH_{\mathrm{eff,D}}$ in the following proposition. 

\begin{proposition}
	The Euclidean gradient $\nabla_{\bH_{\mathrm{eff,D}}} R_{\mathrm{D}}$ is
	\begin{align} \label{grad_R_D_H_eff_D}
		&\nabla_{\bH_{\mathrm{eff,D}}} R_{\mathrm{D}} = \nonumber \\&
		\frac{1}{\ln2\cdot \sigma_{\mathrm{D}}^2} \bH_{\mathrm{eff,D}} \bF_{\mathrm{D}} \left( \bI_{N_{\mathrm{s}}^{\mathrm{D}}} + \frac{1}{\sigma_{\mathrm{D}}^2 } \bF_{\mathrm{D}}^{\mathrm{H}}\bH_{\mathrm{eff,D}} ^{\mathrm{H}}\bH_{\mathrm{eff,D}} \bF_{\mathrm{D}}\right)^{-1} \bF_{\mathrm{D}}^{\mathrm{H}}.
	\end{align}
	\begin{proof}
		See Appendix A.
	\end{proof}
\end{proposition}  

The remaining part in (\ref{Partial_R_D_theta_ell}), i.e., the partial derivative of $\bH_{\mathrm{eff,D}}^{\mathrm{H}}$ with respect to $\theta_{\ell}^*$, can be computed by
\begin{equation} \label{grad_H_eff_D_theta_ell}
	\frac{\partial \bH_{\mathrm{eff,D}}^{\mathrm{H}}}{\partial \theta_{\ell}^*} = (\bh^{\prime}_{\mathrm{D},\ell} \otimes \bG_{\mathrm{D}}(\ell,:))^{\mathrm{H}},
\end{equation}
where $\bh^{\prime}_{\mathrm{D},\ell}$ denotes the $\ell$-th column of $\bH_{\mathrm{D}}^{\mathrm{H}}$.
Based on (\ref{grad_R_D_H_eff_D}) and (\ref{grad_H_eff_D_theta_ell}), the partial derivative of $R_{\mathrm{U}}$ with respect to $\theta_{\ell}$ can be similarly obtained by denoting the $\ell$-th column of $\bG_{\mathrm{U}}^{\mathrm{H}}$ as~$\bg^{\prime}_{\mathrm{U},\ell}$.

\subsubsection{Transport}
Let $\bd^{(t)}$ be the search direction at the point of the $t$-th iteration $\btheta^{(t)}$.
In the manifold optimization, it is likely that the search directions $\bd^{(t)}$ and $\bd^{(t+1)}$ will lie in different tangent spaces, and the transport operation is introduced to address this issue. The transport operation enables mapping the tangent vector from one tangent space to another. In the case of the complex circle manifold $\cM_{cc}^L$, the vector transport is given by
\begin{align} \label{transport}
	\cT_{\btheta^{(t)}\rightarrow \btheta^{(t+1)}}\left(\bd^{(t)}\right) \triangleq \bd^{(t)} -\Re\left(\bd^{(t)} \circ \btheta^{(t+1)*}\right) \circ \btheta^{(t+1)},
\end{align}
which maps the tangent vector $\bd^{(t)}$ at the point $\btheta^{(t)}$ to the tangent space at the point of the next iteration $\btheta^{(t+1)}$.

As the update rule for the search direction after obtaining the Riemannian gradient (\ref{Riemannian_grad}), the conjugate gradient method is used in the RCG algorithm, and the search direction $\bd^{(t+1)}$ is given by
\begin{equation} \label{conjugate_gradient}
	\bd^{(t+1)} = -\operatorname{grad}_{\btheta^{(t+1)}} R_{\mathrm{WSR}} + \gamma^{(t)} \cT_{\btheta^{(t)} \rightarrow \btheta^{(t+1)}}\left(\bd^{(t)}\right),
\end{equation}
where $\gamma^{(t)}$ can be chosen as the Polak-Ribiere parameter \cite{Book:manifold}.

\subsubsection{Retraction}

Let $\tau^{(t)}$ be the step size for the search direction $\bd^{(t)}$ at the point $\btheta^{(t)}$. Note that the step size can be found using the Armijo backtracking line search \cite{Book:manifold}, and the related parameters can be set adaptively. When the point on the manifold moves by $\tau^{(t)} \bd^{(t)}$ along the tangent vector, it may not be on the manifold itself. To tackle this problem, the retraction operation is employed to map the updated point back onto the manifold. Consequently, the updated point $\btheta^{(t+1)}$ can be computed as
\begin{equation} \label{retraction}
	\btheta^{(t+1)} = \cR\left(\btheta^{(t)}+ \tau^{(t)} \bd^{(t)}\right),
\end{equation}
where $\cR(\cdot)$ denotes the retraction operation, which is given~by
\begin{equation}
	\cR(\btheta) = \left[\frac{\theta_1}{\vert \theta_1 \vert},\cdots,\frac{\theta_L}{\vert \theta_L \vert} \right]^{\mathrm{T}}.
\end{equation}
The overall RCG-based algorithm is summarized in Algorithm~1, which is guaranteed to converge to a stationary point for (P2) \cite{Book:manifold}.

Based on Algorithm 1, we summarize the overall algorithm to find the effective solution of (P1) in Algorithm 2. The RIS reflection coefficients are first initialized, and then the precoders at the BS and UE are computed based on this initial value. The estimate of the reflection coefficients is then updated by Algorithm 1, and the precoders are obtained accordingly. The algorithm terminates if the increment in the objective function (\ref{objective_P1}) is less than some predetermined threshold $\epsilon$.

\begin{algorithm}[t]
	\begin{algorithmic} [1]
		\caption{RCG-based Algorithm for Problem (P2)}
		\State \textbf{Initialization:} $\btheta^{(0)}$, $\bd^{(0)}=-\operatorname{grad}_{\btheta^{(0)}} R_{\mathrm{WSR}}$, and set $t~=~0$
		\Repeat
		\State Choose the Armijo backtracking line search step size $\tau^{(t)}$
		\State Find the next point $\btheta^{(t+1)}$ using the retraction in (\ref{retraction})
		\State Compute the Euclidean gradient $\nabla_{\btheta^{(t+1)}} R_{\mathrm{WSR}}$ according to (\ref{Partial_R_D_theta_ell})
		\State Compute the Riemannian gradient $\operatorname{grad}_{\btheta^{(t+1)}}R_{\mathrm{WSR}}$ according to (\ref{Riemannian_grad})
		\State Compute the transport of $\bd^{(t)}$ according to (\ref{transport})
		\State Calculate the conjugate direction $\bd^{(t+1)}$ according to (\ref{conjugate_gradient})
		\State $t \leftarrow t+1$
		\Until $\Vert \operatorname{grad}_{\btheta^{(t)}}R_{\mathrm{WSR}} \Vert_2 \leq \epsilon$
		\State \textbf{Output:} $\btheta^{\star} = \btheta^{(t)}$
	\end{algorithmic}
\end{algorithm} 

\begin{algorithm}[t]
	\begin{algorithmic} [1]
		\caption{Proposed Manifold Optimization-based Algorithm for Problem (P1)}
		\State \textbf{Input:} $\bG_{\mathrm{D}}, \bH_{\mathrm{D}}, \bG_{\mathrm{U}}, \bH_{\mathrm{U}}, \sigma_{\mathrm{D}}, \sigma_{\mathrm{U}}$. 
		\State \textbf{Initialization:} Set $s=0$, and randomly generate $\btheta^{(0)}$, and obtain the optimal $\bF_{\mathrm{D}}^{(0)}$ and $\bF_{\mathrm{U}}^{(0)}$ for the channel realization
		\Repeat
		\State Compute $\btheta^{(s+1)}$ based on \textbf{Algorithm 1} with fixed $\bF_{\mathrm{D}}^{(s)}$ and $\bF_{\mathrm{U}}^{(s)}$
		\State Compute $\bF_{\mathrm{D}}^{(s+1)}$ and $\bF_{\mathrm{U}}^{(s+1)}$ according to (\ref{F_D_opt}) and (\ref{F_U_opt}) with fixed $\btheta^{(s+1)}$
		\State $s \leftarrow s+1$
		\Until $\Vert R_{\mathrm{WSR}}^{(s+1)}-R_{\mathrm{WSR}}^{(s)} \Vert_2 \leq \epsilon$
		\State \textbf{Output:} $\btheta^{\star} = \btheta^{(s)}, \bF_{\mathrm{D}}^{\star}= \bF_{\mathrm{D}}^{(s)}, \bF_{\mathrm{U}}^{\star} = \bF_{\mathrm{U}}^{(s)}$
	\end{algorithmic}
\end{algorithm} 

\subsection{Low-complexity AO}\label{sec4_2}

In this subsection, we propose a low-complexity AO technique for solving 
(P2), where sub-problem (P2) with respect to the $\ell$-th reflection coefficient $\theta_{\ell}$ is formulated with all other reflection coefficients fixed. A closed-form solution for each sub-problem is derived, from which the effective solution of (P2) can be obtained by iteratively solving these sub-problems.

Denote $\bG_{\mathrm{D}}^{\prime} = \bG_{\mathrm{D}} \bF_{\mathrm{D}}=[\bg^{\prime}_{\mathrm{D},1},\cdots,\bg^{\prime}_{\mathrm{D},L}]^{\mathrm{H}}$ $\in \mathbb{C}^{L \times N_{\mathrm{s}}^{\mathrm{D}}}$ and $\bH_{\mathrm{U}}^{\prime} = \bH_{\mathrm{U}} \bF_{\mathrm{U}}=[\bh^{\prime}_{\mathrm{U},1},\cdots,\bh^{\prime}_{\mathrm{U},L}]^{\mathrm{H}}$ $\in \mathbb{C}^{L \times N_{\mathrm{s}}^{\mathrm{U}}}$. Taking a procedure similar to that in \cite{Zhang:2020}, the objective function (\ref{Phase_shift_objective}) can be rewritten with respect to $\theta_{\ell}$ given $\{\theta_{i}, i\neq \ell\}_{i=1}^{L}$, i.e., $f_{\ell}(\theta_{\ell})=\eta f_{\mathrm{D},\ell}(\theta_{\ell}) + (1-\eta)f_{\mathrm{U},\ell} (\theta_{\ell}), \enspace \ell=1,\cdots,L$, where $f_{\mathrm{D},\ell}(\theta_{\ell})$ and $f_{\mathrm{U},\ell} (\theta_{\ell})$ are expressed as
\begin{align} \label{f_ell}
	f_{\mathrm{D},\ell}(\theta_{\ell})&= \log_2 \det\left(\bI_{K} + \theta_{\ell} \bA_{\mathrm{D},\ell}^{-1} \bB_{\mathrm{D},\ell}+\theta_{\ell}^* \bA_{\mathrm{D},\ell}^{-1} \bB_{\mathrm{D},\ell}^{\mathrm{H}}\right) \nonumber \\ & \enspace \enspace \enspace +\log_2 \det(\bA_{\mathrm{D},\ell}) \nonumber \\& \triangleq f^{\prime}_{\mathrm{D},\ell}(\theta_{\ell})+\log_2 \det(\bA_{\mathrm{D},\ell}), \nonumber \\
	f_{\mathrm{U},\ell}(\theta_{\ell}) &= \log_2 \det \left(\bI_{N} + \theta_{\ell} \bA_{\mathrm{U},\ell}^{-1} \bB_{\mathrm{U},\ell}+\theta_{\ell}^* \bA_{\mathrm{U},\ell}^{-1} \bB_{\mathrm{U},\ell}^{\mathrm{H}} \right) \nonumber \\ & \enspace \enspace \enspace +\log_2 \det(\bA_{\mathrm{U},\ell}) \nonumber \\& \triangleq f^{\prime}_{\mathrm{U},\ell}(\theta_{\ell})+\log_2 \det(\bA_{\mathrm{U},\ell}),
\end{align}
with $\bA_{\mathrm{D},\ell} \in \mathbb{C}^{K \times K}$, $\bA_{\mathrm{U},\ell} \in{ \mathbb{C}^{N \times N}}$, $\bB_{\mathrm{D},\ell} \in \mathbb{C}^{K \times K}$, and $\bB_{\mathrm{U},\ell} \in \mathbb{C}^{N \times N}$ given by (\ref{A_B}) at \textcolor{black}{the top of the next page.}
\begin{figure*}[t]
	\begin{align} \label{A_B}
		&\bA_{\mathrm{D},\ell} = \bI_{K} + \frac{1}{\sigma_{\mathrm{D}}^2} \left(  \sum_{i=1,i\neq \ell}^L \theta_i \bh^{\prime}_{\mathrm{D},i} (\bg^{\prime}_{\mathrm{D},i})^{\mathrm{H}} \right)  \left(  \sum_{i=1,i\neq \ell}^L \theta_i \bh^{\prime}_{\mathrm{D},i} (\bg^{\prime}_{\mathrm{D},i})^{\mathrm{H}}  \right)^{\mathrm{H}} + \frac{1}{\sigma_{\mathrm{D}}^2} \bh^{\prime}_{\mathrm{D},\ell}(\bg^{\prime}_{\mathrm{D},\ell})^{\mathrm{H}}\bg^{\prime}_{\mathrm{D},\ell} (\bh^{\prime}_{\mathrm{D},\ell})^{\mathrm{H}}, \nonumber \\
		&\bA_{\mathrm{U},\ell} = \bI_{N} + \frac{1}{\sigma_{\mathrm{U}}^2} \left(  \sum_{i=1,i\neq \ell}^L \theta_i \bg^{\prime}_{\mathrm{U},i} (\bh^{\prime}_{\mathrm{U},i})^{\mathrm{H}} \right)  \left(  \sum_{i=1,i\neq \ell}^L \theta_i \bg^{\prime}_{\mathrm{U},i} (\bh^{\prime}_{\mathrm{U},i})^{\mathrm{H}}  \right)^{\mathrm{H}} + \frac{1}{\sigma_{\mathrm{U}}^2} \bg^{\prime}_{\mathrm{U},\ell}(\bh^{\prime}_{\mathrm{U},\ell})^{\mathrm{H}}\bh^{\prime}_{\mathrm{U},\ell} (\bg^{\prime}_{\mathrm{U},\ell})^{\mathrm{H}}, \nonumber \\
		&\bB_{\mathrm{D},\ell} =  \frac{1}{\sigma_{\mathrm{D}}^2} \bh^{\prime}_{\mathrm{D},\ell}(\bg^{\prime}_{\mathrm{D},\ell})^{\mathrm{H}} \left( \sum_{i=1,i\neq \ell}^L \theta_i^* \bg^{\prime}_{\mathrm{D},i} (\bh_{\mathrm{D},i}^{\prime})^{\mathrm{H}} \right), \nonumber \\ &\bB_{\mathrm{U},\ell} =  \frac{1}{\sigma_{\mathrm{U}}^2} \bg^{\prime}_{\mathrm{U},\ell}(\bh^{\prime}_{\mathrm{U},\ell})^{\mathrm{H}}  \left( \sum_{i=1,i\neq \ell}^L \theta_i^* \bh^{\prime}_{\mathrm{U},i} (\bg_{\mathrm{U},i}^{\prime})^{\mathrm{H}} \right).
	\end{align}
	\hrule
\end{figure*}
Since $f^{\prime}_{\mathrm{D},\ell}(\theta_{\ell})$ and $f^{\prime}_{\mathrm{U},\ell}(\theta_{\ell})$ only depend on $\theta_{\ell}$, the equivalent problem of (P2) with respect to $\theta_{\ell}$ can be formulated as
\begin{align}
	\mbox{(P3)}: \enspace \max_{\theta_{\ell}} \enspace &f^{\prime}_{\ell}(\theta_{\ell})=\eta f^{\prime}_{\mathrm{D},\ell}(\theta_{\ell}) + (1-\eta)f^{\prime}_{\mathrm{U},{\ell}} (\theta_{\ell}) \label{P3_objective} \\ \mbox{s.t.} \enspace \enspace & \vert \theta_{\ell} \vert = 1.
\end{align}

To proceed, it is necessary to investigate properties related to $\bA_{\mathrm{D},\ell}^{-1} \bB_{\mathrm{D},\ell}$ and $\bA_{\mathrm{U},\ell}^{-1} \bB_{\mathrm{U},\ell}$ since the objective function (\ref{P3_objective}) is largely affected by these matrices\footnote{Although there exist additional matrices $\bA_{\mathrm{D},\ell}^{-1} \bB_{\mathrm{D},\ell}^{\mathrm{H}}$ and $\bA_{\mathrm{U},\ell}^{-1} \bB_{\mathrm{U},\ell}^{\mathrm{H}}$ in (\ref{P3_objective}), only the properties of $\bA_{\mathrm{D},\ell}^{-1}\bB_{\mathrm{D},\ell}$ and $\bA_{\mathrm{U},\ell}^{-1} \bB_{\mathrm{U},\ell}$ are required for the reformulation of $f^{\prime}_{\mathrm{D},\ell}(\theta_{\ell})$ and $f^{\prime}_{\mathrm{U},\ell}(\theta_{\ell})$. For instance, this can be checked by following the procedure in Appendix B.}.
Note that the rank of both $\bA_{\mathrm{D},\ell}^{-1} \bB_{\mathrm{D},\ell}$ and $\bA_{\mathrm{U},\ell}^{-1} \bB_{\mathrm{U},\ell}$ is upper bounded by 1 since $\mathrm{rank}\left(\bB_{\mathrm{D},\ell}\right)=\mathrm{rank}\left(\bB_{\mathrm{U},\ell}\right)=1$, and if the rank of either of these matrices is zero, the corresponding parts of (\ref{P3_objective}) become independent of $\theta_{\ell}$. For instance, if the rank of $\bA_{\mathrm{D},\ell}^{-1} \bB_{\mathrm{D},\ell}$ is zero, i.e., $\bA_{\mathrm{D},\ell}^{-1} \bB_{\mathrm{D},\ell}= \b0_{K,K}$, 
$f^{\prime}_{\mathrm{D},\ell}(\theta_{\ell})$ is independent of $\theta_{\ell}$, and the solution to (P3) can be obtained by maximizing $f^{\prime}_{\mathrm{U},\ell}(\theta_{\ell})$ only.

When both matrices are rank-one, two sub-cases are possible depending on whether or not the matrices are diagonalizable. 
According to \cite{Zhang:2020}, it can be shown that diagonalizable $\bA_{\mathrm{D},\ell}^{-1} \bB_{\mathrm{D},\ell}$ and $\bA_{\mathrm{U},\ell}^{-1} \bB_{\mathrm{U},\ell}$ are equivalent to $\trace\left(\bA_{\mathrm{D},\ell}^{-1} \bB_{\mathrm{D},\ell} \right) \neq 0$ and $\trace\left(\bA_{\mathrm{U},\ell}^{-1} \bB_{\mathrm{U},\ell} \right) \neq 0$.

\subsubsection{Diagonalizable $\bA_{\mathrm{D},\ell}^{-1} \bB_{\mathrm{D},\ell}$ and $\bA_{\mathrm{U},\ell}^{-1} \bB_{\mathrm{U},\ell}$}
In this case, both matrices can be decomposed by an eigenvalue decomposition (EVD):
\begin{align} \label{EVD_AB}
	\bA_{\mathrm{D},\ell}^{-1} \bB_{\mathrm{D},\ell} &=\bU_{\mathrm{D},\ell} 
	\bLambda_{\mathrm{D},\ell} \bU_{\mathrm{D},\ell}^{-1}, \nonumber \\
	\bA_{\mathrm{U},\ell}^{-1} \bB_{\mathrm{U},\ell} &=\bU_{\mathrm{U},\ell} \bLambda_{\mathrm{U},\ell} \bU_{\mathrm{U},\ell}^{-1},
\end{align}
where $\bLambda_{\mathrm{D},\ell}=\diag(\lambda_{\mathrm{D},\ell},0,\cdots,0)$ and $\bLambda_{\mathrm{U},\ell}=\diag(\lambda_{\mathrm{U},\ell},0,\cdots,0)$, each with only a single non-zero eigenvalue $\lambda_{\mathrm{D},\ell}$ and~$\lambda_{\mathrm{U},\ell}$, respectively.
Before proceeding further, let $(\bc_{\mathrm{D},\ell}^{\prime})^{\mathrm{T}}$ and $(\bc_{\mathrm{U},\ell}^{\prime})^{\mathrm{T}}$ denote the first rows of $\bC_{\mathrm{D},\ell}= \bU_{\mathrm{D},\ell}^{\mathrm{H}} \bA_{\mathrm{D},\ell} \bU_{\mathrm{D},\ell}$ and $\bC_{\mathrm{U},\ell}= \bU_{\mathrm{U},\ell}^{\mathrm{H}} \bA_{\mathrm{U},\ell} \bU_{\mathrm{U},\ell}$, and let $\bc_{\mathrm{D},\ell}$ and $\bc_{\mathrm{U},\ell}$ denote the first columns of $\bC_{\mathrm{D},\ell}^{-1}$ and $\bC_{\mathrm{U},\ell}^{-1}$, respectively.
Following \cite{Zhang:2020} and using some mathematical manipulations, $f^{\prime}_{\mathrm{D},\ell}(\theta_{\ell})$ and $f^{\prime}_{\mathrm{U},\ell}(\theta_{\ell})$ can be rewritten as
\begin{align} \label{f_prime_ell}
	f^{\prime}_{\mathrm{D},\ell}(\theta_{\ell}) &= \log_2(1+\vert \lambda_{\mathrm{D},\ell} \vert^2 (1-c^{\prime}_{\mathrm{D},\ell1} c_{\mathrm{D},\ell1})+2\mathrm{Re}(\theta_{\ell} \lambda_{\mathrm{D},\ell})), \nonumber \\
	f^{\prime}_{\mathrm{U},\ell}(\theta_{\ell}) &= \log_2(1+\vert \lambda_{\mathrm{U},\ell} \vert^2 (1-c^{\prime}_{\mathrm{U},\ell1} c_{\mathrm{U},\ell1})+2\mathrm{Re}(\theta_{\ell} \lambda_{\mathrm{U},\ell})),
\end{align}
where $c^{\prime}_{\mathrm{D},\ell1}$ and $c^{\prime}_{\mathrm{U},\ell1}$ are the first elements of $(\bc_{\mathrm{D},\ell}^{\prime})^{\mathrm{T}}$ and $(\bc_{\mathrm{U},\ell}^{\prime})^{\mathrm{T}}$, and $c_{\mathrm{D},\ell1}$ and $c_{\mathrm{U},\ell1}$ are the first elements of $\bc_{\mathrm{D},\ell}$ and $\bc_{\mathrm{U},\ell}$, respectively.
The detailed procedure to derive (\ref{f_prime_ell}) is provided in Appendix B.

Denote the singular values of an arbitrary $n \times n$ complex matrix $\bA$ by $\rho_1(\bA)\geq\cdots\geq \rho_n(\bA)$. 
To exploit the structure of (\ref{f_prime_ell}), we provide the following lemma to identify an upper bound for~$\vert \lambda_{\mathrm{D},\ell} \vert$.
\begin{lemma}
	The quantity $\vert \lambda_{\mathrm{D},\ell} \vert$ is upper bounded by $\rho_1\left( \bA_{\mathrm{D},\ell}^{-1} \right) \rho_1\left( \bB_{\mathrm{D},\ell} \right)$.
	\begin{proof}
		Since $\bA_{\mathrm{D},\ell}^{-1} \bB_{\mathrm{D},\ell}$ is a rank-one matrix, an upper bound for $\vert \lambda_{\mathrm{D},\ell} \vert$ can be found as follows:
		\begin{align} \label{lambda_D_bound}
			\vert \lambda_{\mathrm{D},\ell} \vert &= \left\vert \trace\left( \bA_{\mathrm{D},\ell}^{-1} \bB_{\mathrm{D},\ell} \right) \right\vert \nonumber \\& \mathop \leq \limits^{(a)} \sum_{k=1}^K \rho_k\left( \bA_{\mathrm{D},\ell}^{-1} \right) \rho_k\left( \bB_{\mathrm{D},\ell} \right) \nonumber \\ &\mathop = \limits^{(b)} \rho_1\left( \bA_{\mathrm{D},\ell}^{-1} \right) \rho_1\left( \bB_{\mathrm{D},\ell} \right),
		\end{align}
		where $(a)$ follows from Von Neumann's trace inequality \cite{Trace_inequality}, and $(b)$ holds since $\bB_{\mathrm{D},\ell}$ is rank-one.
	\end{proof}
\end{lemma}

Note that $\rho_1\left( \bA_{\mathrm{D},\ell}^{-1}\right)$ is equivalent to the smallest singular value of $\bA_{\mathrm{D},\ell}$, i.e., $\rho_K \left(\bA_{\mathrm{D},\ell} \right)$, since $\bA_{\mathrm{D},\ell}$ is symmetric. To further investigate the upper bound of $\vert \lambda_{\mathrm{D},\ell} \vert$, we provide the following lemma related to the smallest singular value of $\bA_{\mathrm{D},\ell}$.
\begin{lemma} \label{lemma_2}
	If $N_{\mathrm{s}}^{\mathrm{D}}<K-1$, the smallest singular value of $\bA_{\mathrm{D},\ell}$ is given by $\rho_K\left( \bA_{\mathrm{D},\ell} \right)=1$.
	\begin{proof}
		Define $\bX=\sum_{i=1,i\neq \ell}^L \theta_i \bh^{\prime}_{\mathrm{D},i} (\bg^{\prime}_{\mathrm{D},i})^{\mathrm{H}}$ and $\bY=\bh^{\prime}_{\mathrm{D},\ell}(\bg^{\prime}_{\mathrm{D},\ell})^{\mathrm{H}}$, which form $\bA_{\mathrm{D},\ell}$ as in (\ref{A_B}). The maximum ranks of $\bX'=\bX\bX^{\mathrm{H}}$ and $\bY'=\bY\bY^{\mathrm{H}}$ are given by $\mathrm{rank}\left(\bX' \right)=N_{\mathrm{s}}^{\mathrm{D}}$ and $\mathrm{rank}\left(\bY' \right)=1$, and we consider these maximum values to obtain the tightest bound for $N_{\mathrm{s}}^{\mathrm{D}}$.
		Let the eigenvalues of an arbitrary $n \times n$ Hermitian matrix $\bA$ be $\lambda_1(\bA)\geq \cdots \geq \lambda_n(\bA)$.
		According to Weyl's inequality \cite{matrix_analysis}, an upper bound for $\lambda_K\left( \bX' + \bY' \right)$ is given by
		\begin{align} \label{Weyl_inequality}
			\lambda_K\left(\bX'+\bY'\right) \leq \lambda_i\left(\bX'\right) + \lambda_{K+1-i}\left(\bY'\right), \enspace i=1,\cdots,K.
		\end{align}
		When $i=N_{\mathrm{s}}^{\mathrm{D}}+1$, (\ref{Weyl_inequality}) can be simplified to $\lambda_K\left(\bX'+\bY'\right) \leq 0$ when $N_{\mathrm{s}}^{\mathrm{D}}<K-1$, resulting in $\lambda_K\left(\bX'+\bY'\right)=0$ due to the positive semi-definiteness of $\bX'+\bY'$, which implies that $\lambda_K\left( \bA_{\mathrm{D},\ell} \right)$ is only affected by $\bI_K$ as clearly shown in (\ref{A_B}).
		Hence, $\lambda_K\left( \bA_{\mathrm{D},\ell} \right)=1$, and thereby $\rho_K\left( \bA_{\mathrm{D},\ell} \right)=1$, which completes the proof.
	\end{proof}
\end{lemma}
From Lemma \ref{lemma_2}, the upper bound (\ref{lambda_D_bound}) can be simplified to $\rho_1\left( \bB_{\mathrm{D},\ell} \right)$, which is equivalent to $\left\Vert  \bB_{\mathrm{D},\ell} \right\Vert_{\mathrm{F}}$ because the other singular values of $\bB_{\mathrm{D},\ell}$ are zero.
Similarly, it can be shown that $\vert \lambda_{\mathrm{U},\ell} \vert$ is upper bounded by $\rho_1\left( \bA_{\mathrm{U},\ell}^{-1} \right) \rho_1\left( \bB_{\mathrm{U},\ell} \right)$, which can be simplified to $\Vert \bB_{\mathrm{U},\ell} \Vert_{\mathrm{F}}$ when $N_{\mathrm{s}}^{\mathrm{U}}<N-1$.

Based on the above discussion, $\vert \lambda_{\mathrm{D},\ell} \vert$ and $\vert \lambda_{\mathrm{U},\ell} \vert$ are upper bounded by $\left\Vert \bB_{\mathrm{D},\ell} \right\Vert_{\mathrm{F}}$ and $\left\Vert \bB_{\mathrm{U},\ell} \right\Vert_{\mathrm{F}}$ under the respective conditions $N_{\mathrm{s}}^{\mathrm{D}}<K-1$ and $N_{\mathrm{s}}^{\mathrm{U}}<N-1$. According to (\ref{A_B}), $\left\Vert \bB_{\mathrm{D},\ell} \right\Vert_{\mathrm{F}}$ and $\left\Vert \bB_{\mathrm{U},\ell} \right\Vert_{\mathrm{F}}$ are strongly affected by the gains of the RIS-related channels. Note that the RIS will usually be deployed to extend signal coverage and support weak or distant users, implying that $\vert \lambda_{\mathrm{D},\ell} \vert$ and $\vert \lambda_{\mathrm{U},\ell} \vert$ can be assumed to be sufficiently small.
For this reason, we apply the first-order Taylor approximation $\log(1+x) \approx x$ around $x=0$ in (\ref{f_prime_ell}) which, after removing irrelevant terms involving $\theta_{\ell}$, leads to the following approximation of problem (P3):
\begin{align} \label{P3}
	\mbox{(P4)}: \enspace \max_{\theta_{\ell}} \enspace &f^{\prime \prime}_{\ell}(\theta_{\ell}) = \eta\mathrm{Re}(\theta_{\ell} \lambda_{\mathrm{D},\ell}) + (1-\eta)\mathrm{Re}(\theta_{\ell} \lambda_{\mathrm{U},\ell}) \nonumber \\ \mbox{ s.t. } \enspace &\vert \theta_{\ell} \vert = 1.
\end{align}
In the following proposition, we derive the closed-form solution to (P4).

\begin{proposition}
	The optimal solution to (P4) is given by
	\begin{equation} \label{opt_theta_ell}
		\theta_{\ell}^{\star} = \exp(-j\arg(\phi_{\ell})),
	\end{equation}
	where $\phi_{\ell}$ is given in (\ref{phi_ell}) at the \textcolor{black}{top of this page.}
	
	\begin{proof}
		The objective function of (P4) can be rewritten by the weighted sum of cosine functions 
		\begin{align}
			f^{\prime \prime}_{\ell}(\theta_{\ell})  &= \eta \vert \lambda_{\mathrm{D},\ell} \vert \cos(\arg(\theta_{\ell})+\arg(\lambda_{\mathrm{D},\ell}))\nonumber \\& \enspace \enspace +(1-\eta) \vert \lambda_{\mathrm{U},\ell} \vert \cos(\arg(\theta_{\ell})+\arg(\lambda_{\mathrm{U},\ell})) \label{Proposition2_1} \\&= A\cos(\arg(\theta_{\ell})+\phi_{\ell}).\label{Proposition2_2}
		\end{align}
		Applying the trigonometric identity $\cos(\alpha+\beta)=\cos\alpha \cos\beta - \sin\alpha \sin\beta$ to (\ref{Proposition2_1}) and (\ref{Proposition2_2}) yields
		\begin{align}
			A\cos(\phi_{\ell})= &\eta \vert \lambda_{\mathrm{D},\ell}\vert \cos(\arg(\lambda_{\mathrm{D},\ell})) \nonumber \\&+(1-\eta)\vert \lambda_{\mathrm{U},\ell}\vert \cos(\arg(\lambda_{\mathrm{U},\ell})), \label{Proposition2_3} \\  A\sin(\phi_{\ell})= & \eta \vert \lambda_{\mathrm{D},\ell}\vert \sin(\arg(\lambda_{\mathrm{D},\ell})) \nonumber \\&+(1-\eta)\vert \lambda_{\mathrm{U},\ell}\vert \sin(\arg(\lambda_{\mathrm{U},\ell})). \label{Proposition2_4}
		\end{align}
		From (\ref{Proposition2_3}) and (\ref{Proposition2_4}), $\tan(\phi_{\ell})$ can be derived, and it can be shown that $\phi_{\ell}$ is equivalent to (\ref{phi_ell}). Hence, the maximum value of (\ref{Proposition2_2}) can be achieved when $\theta_{\ell}$ is given as (\ref{opt_theta_ell}), which completes the proof. Note that the amplitude $A$ in (\ref{Proposition2_2}) can be directly computed based on (\ref{Proposition2_3}) and (\ref{Proposition2_4}), and this solution is substituted into the original objective function for (P3). 
	\end{proof}
\end{proposition}

\begin{figure*}[t!] 
	\begin{align} \label{phi_ell}
		\phi_{\ell} = \tan^{-1}\left( \frac{\eta \vert \lambda_{\mathrm{D},\ell} \vert \sin(\arg(\lambda_{\mathrm{D},\ell})) + (1-\eta)\vert \lambda_{\mathrm{U},\ell} \vert\sin(\arg(\lambda_{\mathrm{U},\ell}))}{\eta \vert \lambda_{\mathrm{D},\ell} \vert \cos(\arg(\lambda_{\mathrm{D},\ell})) + (1-\eta)\vert \lambda_{\mathrm{U},\ell} \vert\cos(\arg(\lambda_{\mathrm{U},\ell}))  } \right),
	\end{align}
\end{figure*}

\subsubsection{Non-diagonalizable $\bA_{\mathrm{D},\ell}^{-1} \bB_{\mathrm{D},\ell}$ or $\bA_{\mathrm{U},\ell}^{-1} \bB_{\mathrm{U},\ell}$}

According to \cite{Zhang:2020}, it can be verified that if $\bA_{\mathrm{D},\ell}^{-1} \bB_{\mathrm{D},\ell}$ or  $\bA_{\mathrm{U},\ell}^{-1} \bB_{\mathrm{U},\ell}$ is non-diagonalizable, either $f^{\prime}_{\mathrm{D},\ell}(\theta_{\ell})$ or $f^{\prime}_{\mathrm{U},\ell}(\theta_{\ell})$ becomes independent of $\theta_{\ell}$, and the solution to (P3) can be obtained by maximizing only the diagonalizable part,
i.e., $\theta_{\ell}^{\star}=\exp(-j\arg(\lambda_{\mathrm{U},\ell}))$ or $\exp(-j\arg(\lambda_{\mathrm{D},\ell}))$. If both matrices are non-diagonalizable, any choice of reflection coefficients is optimal, e.g., $\theta_{\ell}^{\star}=1$ without loss of generality.

\subsubsection{Summary}
The proposed low-complexity AO technique for solving (P2) is summarized in (\ref{theta_opt}) at the \textcolor{black}{top of the next page}, and the overall low-complexity AO-based algorithm for (P1) is described in Algorithm 3.
The difference compared to the manifold optimization-based algorithm lies in the optimization of the reflection coefficients, which are updated individually using the derived closed-form expressions.

\textit{Remark 1:} In the manifold optimization-based algorithm, monotonic convergence is guaranteed since in every sub-problem the objective function (\ref{objective_P1}) is monotonically increasing, and (\ref{objective_P1}) is upper bounded by finite channel capacities corresponding to the downlink and uplink channels.
Hence, given that (\ref{objective_P1}) is differentiable, the manifold optimization-based algorithm is guaranteed to converge to a stationary point \cite{Convergence}. In the low-complexity AO approach, the closed-form solution (\ref{opt_theta_ell}) is derived from the approximated problem (P4), which in theory eliminates the convergence guarantee. However, the update procedures for the downlink and uplink precoders always maximize their corresponding sub-problems, implying that at least a point near local optima can be obtained. The convergence behavior of the proposed algorithms will be discussed in Section \ref{sec5_2}.

\subsection{Generalized scenarios}
Throughout this paper, we have focused on a scenario involving a single BS communicating with a single UE in FDD. However, our proposed algorithms can be easily extended to the following more general scenarios:
\begin{itemize}
\item \textit{Multiple FDD links where a single BS serves multiple UEs}: In this scenario, the BS communicates with several UEs simultaneously on distinct downlink and uplink frequencies, for example using orthogonal frequency division multiplexing (OFDM). In this case, the objective function~(\ref{objective_P1}) can be extended to incorporate a weighted sum-rate for all UEs, and the proposed algorithms can be directly applied since there is no interference among the different links. The computation of the precoders at the BS and all UEs using eigenmode transmissions would be implemented as before. For the manifold optimization, the Euclidean gradient of the new objective function can be computed by aggregating the individual Euclidean gradients corresponding to each UE, and thus it too can be implemented using the same approach described previously.
Although the number of the closed-form expressions needed for the low-complexity AO approach may increase, these additional terms can be derived in a straightforward way.
\item \textit{Multiple BSs where different service providers exploit distinct frequency bands}:
This scenario is a further generalization of the prior one, and the optimization can be performed similarly. However, in this case the optimization would require sharing global CSI among all entities involved in the optimization, and the RIS would be operated for a common purpose rather than for a specific service provider.
\end{itemize}

\begin{figure*}[t]
	\begin{align} \label{theta_opt}
		\theta_{\ell}^{\star}= \begin{cases} \exp(-j \arg(\phi_{\ell})), &\mbox{ if } \trace\left( \bA_{\mathrm{D},\ell}^{-1} \bB_{\mathrm{D},\ell} \right) \neq 0 \mbox{ and } \trace\left( \bA_{\mathrm{U},\ell}^{-1} \bB_{\mathrm{U},\ell} \right) \neq 0, \\ \exp(-j \arg( \lambda_{\mathrm{D},\ell}) ), &\mbox{ if } \trace\left( \bA_{\mathrm{D},\ell}^{-1} \bB_{\mathrm{D},\ell} \right) \neq 0 \mbox{ and } \trace\left( \bA_{\mathrm{U},\ell}^{-1} \bB_{\mathrm{U},\ell} \right) = 0, \\ \exp(-j \arg( \lambda_{\mathrm{U},\ell})), &\mbox{ if } \trace\left( \bA_{\mathrm{D},\ell}^{-1} \bB_{\mathrm{D},\ell} \right) = 0 \mbox{ and } \trace\left( \bA_{\mathrm{U},\ell}^{-1} \bB_{\mathrm{U},\ell} \right) \neq 0, \\ 1, &\mbox{ otherwise.} \end{cases}
	\end{align}
	\hrule
\end{figure*}

\begin{algorithm}[t]
	\begin{algorithmic} [1]
		\caption{Proposed Low-complexity AO-based Algorithm for Problem (P1)}
		\State \textbf{Input:} $\bG_{\mathrm{D}}, \bH_{\mathrm{D}}, \bG_{\mathrm{U}}, \bH_{\mathrm{U}}, \sigma_{\mathrm{D}}, \sigma_{\mathrm{U}}$
		\State \textbf{Initialization:} Set $t=0$, and randomly generate $\btheta^{(0)}$, and obtain the optimal $\bF_{\mathrm{D}}^{(0)}$ and $\bF_{\mathrm{U}}^{(0)}$ for the channel realization
		\Repeat
		\For{$\ell=1 \rightarrow L$}
		\State Compute $\bA_{\mathrm{D},\ell}, \bA_{\mathrm{U},\ell}, \bB_{\mathrm{D},\ell}$, and $\bB_{\mathrm{U},\ell}$ according to (\ref{A_B})
		\State Obtain $\theta_{\ell}^{(t+1)}$ according to (\ref{theta_opt})
		\EndFor
		\State Compute $\bF_{\mathrm{D}}^{(t+1)}$ and $\bF_{\mathrm{U}}^{(t+1)}$ according to (\ref{F_D_opt}) and (\ref{F_U_opt}) with fixed $\btheta^{(t+1)}$
		\State $t \leftarrow t+1$
		\Until $\Vert R_{\mathrm{WSR}}^{(t+1)}-R_{\mathrm{WSR}}^{(t)} \Vert_2 \leq \epsilon$
		\State \textbf{Output:} $\btheta^{\star} = \btheta^{(t)}, \bF_{\mathrm{D}}^{\star}= \bF_{\mathrm{D}}^{(t)}, \bF_{\mathrm{U}}^{\star} = \bF_{\mathrm{U}}^{(t)}$
	\end{algorithmic}
\end{algorithm} 

\subsection{Complexity analysis}
For simplicity, we assume $(N_{\mathrm{s}}^{\mathrm{D}}, N_{\mathrm{s}}^{\mathrm{U}}) \ll (N,K) \leq L$ to compare the complexity between the proposed~algorithms.
\subsubsection{Manifold optimization-based algorithm}
To optimize $\bF_{\mathrm{D}}$ or $\bF_{\mathrm{U}}$, the worst-case complexity is given by $\cO(NK\min(N,K))$. 
The complexity for implementing the RCG-based algorithm is dominated by computing the Euclidean gradient $\nabla_{\btheta} R_{\mathrm{WSR}}$.
For the downlink part, the required complexities to compute $\nabla_{\bH_{\mathrm{eff,D}}}R_{\mathrm{D}}$ and $\frac{\partial \bH_{\mathrm{eff,D}}^{\mathrm{H}}}{\partial \theta_{\ell}^*}$ are given by $\cO(3NKN_{\mathrm{s}}^{\mathrm{D}})$ and $\cO(NK)$. The complexity for the multiplication between $\nabla_{\bH_{\mathrm{eff,D}}}R_{\mathrm{D}}$ and $\frac{\partial \bH_{\mathrm{eff,D}}^{\mathrm{H}}}{\partial \theta_{\ell}^*}$ is $\cO(N^2K)$. Therefore, the complexity to compute $\nabla_{\btheta} R_{\mathrm{D}}$ is $\cO(N^2KL)$.
Similarly, for the uplink part, the required complexity to compute $\nabla_{\btheta} R_{\mathrm{U}}$ is $\cO( NK^2L )$.
In the retraction operation, the step size $\tau^{(t)}$ should be searched for at each iteration, and the complexity is $O(L)$.
Thus, the total complexity for the manifold optimization-based algorithm can be shown to be $\cO(I_{\mathrm{out},1}I_{\mathrm{in}} (N^2KL + NK^2L))$, where $I_{\mathrm{out},1}$ denotes the number of outer iterations for the entire algorithm until $R_{\mathrm{WSR}}$ converges after all variables are updated, and $I_{\mathrm{in}}$ denotes the number of inner iterations for the RCG-based algorithm given $\bF_{\mathrm{D}}$ and $\bF_{\mathrm{U}}$.

\textit{Remark 2:} In general, the number of outer iterations increases with $L$ accounting for the enlarged search dimension. However, in the manifold optimization-based algorithm, the effect of increasing $L$ is mitigated by the inner iterations to optimize the reflection coefficients.

\subsubsection{Low-complexity AO-based algorithm}
The complexity required to compute $\bG_{\mathrm{D}}^{\prime}$ and $\bH_{\mathrm{U}}^{\prime}$ given $\bF_{\mathrm{D}}$ and $\bF_{\mathrm{U}}$ are $\cO(N^2L)$ and $\cO(K^2L)$, respectively. The worst-case complexity for computing $\lambda_{\mathrm{D},\ell}$ and $\lambda_{\mathrm{U},\ell}$ can be shown to be $\cO(3K^3+2N^2K)$ and $\cO(3N^3+2K^2N)$, respectively. After obtaining $\lambda_{\mathrm{D},\ell}$ and $\lambda_{\mathrm{U},\ell}$, $\theta_{\ell}^{\star}$ are computed by the closed-form expressions which have a negligible complexity. Therefore, the total cost for the low-complexity AO-based algorithm is $\cO( I_{\mathrm{out,2}} (3(N^3+K^3)+2NK(N+K))L )$, where $I_{\mathrm{out},2}$ denotes the number of outer iterations until $R_{\mathrm{WSR}}$ converges.

\subsubsection{Comparison}
Assuming $N=K$, the complexities for the manifold optimization-based and low-complexity AO-based algorithms are given by $\cO(I_{\mathrm{out,1}}I_{\mathrm{in}} N^3L)$ and $\cO( I_{\mathrm{out,2}}N^3L)$, respectively. Due to the additional $I_{\mathrm{in}}$ term required to implement the RCG-based algorithm, the complexity of the manifold optimization is larger than that of the low-complexity AO approach.

\section{Numerical Results}\label{sec5}
\subsection{Simulation scenario} \label{sec5_1}

In this section, we investigate the performance of the proposed algorithms for maximizing the weighted sum-rate for the downlink and uplink transmissions. The BS and UE are assumed to be equipped with a uniform linear array (ULA), while the RIS is modeled as a uniform planar array (UPA) with $L_{\mathrm{h}}$ horizontal rows and $L_{\mathrm{v}}$ vertical columns. We assume $N=16$ BS antennas, $K=8$ UE antennas.
The locations of the BS, RIS, and UE are set to be (0 m, 0 m), (750 m, 5 m), and (800 m, 0 m), respectively. The downlink and uplink carrier frequencies are $f_{\mathrm{D}}=\mbox{2.135 GHz}$ and $f_{\mathrm{U}}=\mbox{1.945 GHz}$.
With the noise spectral density -174 dBm/Hz and the bandwidth 10 MHz, the noise variance is set as $\sigma_{\mathrm{D}}^2 = \sigma_{\mathrm{U}}^2 =\mbox{-104 dBm}$. Unless otherwise specified, the weight coefficient is taken to be $\eta=0.5$, the number of RIS elements is $L=100$ with $L_{\mathrm{h}}=L_{\mathrm{v}}=10$, and the maximum downlink and uplink powers are $P_{\mathrm{D,max}}=\mbox{27 dBm}$ and $P_{\mathrm{U,max}}=\mbox{23 dBm}$.
We further adopt the 3GPP distance and frequency dependent path-loss model given by \cite{Abouamer:2022, 3GPP:path_loss}
\begin{equation}
	\mathrm{PL}\left(d, f\right)\mbox{ [dB]} = 28+22\log_{10}\left(\frac{d}{d_{\mathrm{0}}}\right)+20\log_{10}\left(\frac{f}{f_{\mathrm{0}}}\right),
\end{equation}
where $d$ is the link distance, $f$ is the carrier frequency, and $d_0=\mbox{1 m}$ and $f_0=\mbox{1 GHz}$ respectively denote the reference distance and~frequency.

We adopt the geometric channel model \cite{Shen:2021, Zhou:2023}, which for the downlink BS-RIS channel $\bG_{\mathrm{D}}$ can be written as\footnote{While we assume a geometric channel model for the simulations, the proposed algorithms do not rely on this assumption.}
\begin{align} \label{G_D}
	\bG_{\mathrm{D}}=\sqrt{\frac{N L}{M_{\mathrm{G}}^{\mathrm{D}}}}\sum_{m=1}^{M_{\mathrm{G}}^{\mathrm{D}}}\alpha_{\mathrm{D},m} \ba_{\mathrm{R}}(\delta^{\mathrm{D}}_{\mathrm{G},m}, \gamma^{\mathrm{D}}_{\mathrm{G},m}) (\ba_{\mathrm{B}}(\omega^{\mathrm{D}}_{\mathrm{G},m}))^{\mathrm{H}},
\end{align}
where $M_{\mathrm{G}}^{\mathrm{D}}$ is the number of downlink paths, and $\alpha_{\mathrm{D},m}$ is the downlink channel coefficient of the $m$-th path in $\bG_{\mathrm{D}}$ which we assume to be independent and identically distributed (i.i.d.) as $\alpha_{\mathrm{D},m} \simeq \cC\cN(0, \mathrm{PL}(d_{\mathrm{BR}},f_{\mathrm{D}}))$ for the BS-RIS link distance $d_{\mathrm{BR}}$. The array response vectors at the BS and RIS are respectively denoted by $\ba_{\mathrm{B}}(\omega^{\mathrm{D}}_{\mathrm{G},m}) \in \mathbb{C}^{N \times 1}$ and $\ba_{\mathrm{R}}(\delta^{\mathrm{D}}_{\mathrm{G},m}, \gamma^{\mathrm{D}}_{\mathrm{G},m}) \in \mathbb{C}^{L\times 1}$, and $\ba_{\mathrm{B}}(\omega^{\mathrm{D}}_{\mathrm{G},m})$ is given by
\begin{align} \label{array_response_BS_D}
	\ba_{\mathrm{B}}(\omega^{\mathrm{D}}_{\mathrm{G},m}) = \frac{1}{\sqrt{N}}\left[1,e^{j\omega^{\mathrm{D}}_{\mathrm{G},m}},\cdots,e^{j(N-1)\omega^{\mathrm{D}}_{\mathrm{G},m}} \right]^{\mathrm{T}},
\end{align}
where $\omega^{\mathrm{D}}_{\mathrm{G},m}=2\pi f_{\mathrm{D}} a_{\mathrm{B}} \sin(\zeta_{\mathrm{G},m}^{\mathrm{D}}) / c$ is the spatial frequency of the angle of departure (AoD) assuming the speed of light is $c$, the antenna spacing at the BS is $a_{\mathrm{B}}$, and the AoD is $\zeta_{\mathrm{G},m}^{\mathrm{D}}$. 
The array response vector at the RIS $\ba_{\mathrm{R}}(\delta^{\mathrm{D}}_{\mathrm{G},m}, \gamma^{\mathrm{D}}_{\mathrm{G},m})$ can be written as
\begin{equation}  \label{array_response_RIS_D}
	\ba_{\mathrm{R}}(\delta^{\mathrm{D}}_{\mathrm{G},m}, \gamma^{\mathrm{D}}_{\mathrm{G},m}) = \ba_{\mathrm{R,v}}(\gamma^{\mathrm{D}}_{\mathrm{G},m}) \otimes \ba_{\mathrm{R,h}}(\delta^{\mathrm{D}}_{\mathrm{G},m}), 
\end{equation}
where $\ba_{\mathrm{R,v}}(\gamma^{\mathrm{D}}_{\mathrm{G},m})$ and $\ba_{\mathrm{R,h}}(\delta^{\mathrm{D}}_{\mathrm{G},m})$ represent the RIS array response vectors along the vertical and horizontal directions, and $\ba_{\mathrm{R,v}}(\gamma^{\mathrm{D}}_{\mathrm{G},m})$ and $\ba_{\mathrm{R,h}}(\delta^{\mathrm{D}}_{\mathrm{G},m})$ are given by
\begin{align}
	&\ba_{\mathrm{R,v}}(\gamma^{\mathrm{D}}_{\mathrm{G},m}) =\frac{1}{\sqrt{L_{\mathrm{v}}}}\left[1,e^{j \gamma^{\mathrm{D}}_{\mathrm{G},m} },\cdots, e^{j(L_{\mathrm{v}}-1)\gamma^{\mathrm{D}}_{\mathrm{G},m}} \right]^{\mathrm{T}}, \nonumber
	\\& \ba_{\mathrm{R,h}}(\delta^{\mathrm{D}}_{\mathrm{G},m}) = \frac{1}{\sqrt{L_{\mathrm{h}}}}\left[1,e^{j \delta^{\mathrm{D}}_{\mathrm{G},m} },\cdots, e^{j(L_{\mathrm{h}}-1)\delta^{\mathrm{D}}_{\mathrm{G},m}} \right]^{\mathrm{T}},
\end{align}  
where $\gamma^{\mathrm{D}}_{\mathrm{G},m}= 2\pi f_{\mathrm{D}} a_{\mathrm{R,v}} \sin(\psi_{\mathrm{G},m}^{\mathrm{D}})/ c$ and $\delta^{\mathrm{D}}_{\mathrm{G},m}= 2\pi f_{\mathrm{D}} a_{\mathrm{R,h}} \cos(\psi_{\mathrm{G},m}^{\mathrm{D}}) \sin(\phi_{\mathrm{G},m}^{\mathrm{D}}) / c$ are the spatial frequencies for the angles of arrival (AoAs), the vertical and horizontal spacing of the RIS elements are $a_{\mathrm{R,v}}$ and $a_{\mathrm{R,h}}$, respectively, and the azimuth and elevation AoAs are $\phi_{\mathrm{G},m}^{\mathrm{D}}$ and $\psi_{\mathrm{G},m}^{\mathrm{D}}$, respectively.

Similarly, the uplink RIS-BS channel $\bG_{\mathrm{U}}^{\mathrm{H}}$ can be represented by
\begin{align} \label{G_U}
    \bG_{\mathrm{U}}^{\mathrm{H}}=\sqrt{\frac{N L}{M_{\mathrm{G}}^{\mathrm{U}}}}\sum_{m=1}^{M_{\mathrm{G}}^{\mathrm{U}}} \alpha_{\mathrm{U}, m } \ba_{\mathrm{B}}(\omega^{\mathrm{U}}_{\mathrm{G},m})(\ba_{\mathrm{R}}(\delta^{\mathrm{U}}_{\mathrm{G},m}, \gamma^{\mathrm{U}}_{\mathrm{G},m}))^{\mathrm{H}},
\end{align}
where $M_{\mathrm{G}}^{\mathrm{U}}$ is the number of uplink paths, $\alpha_{\mathrm{U},m}  \simeq \cC \cN(0, \mathrm{PL}(d_{\mathrm{BR}}, f_{\mathrm{U}}))$ is the i.i.d. uplink channel coefficient of the $m$-th path, $\omega^{\mathrm{U}}_{\mathrm{G},m}=2\pi f_{\mathrm{U}} a_{\mathrm{B}} \sin(\zeta_{\mathrm{G},m}^{\mathrm{U}}) / c$ is the AoA spatial frequency at the BS for AoA $\zeta_{\mathrm{G},m}^{\mathrm{U}}$, and $\gamma^{\mathrm{U}}_{\mathrm{G}, m}= 2\pi f_{\mathrm{U}} a_{\mathrm{R,v}} \sin(\psi_{\mathrm{G},m}^{\mathrm{U}})/ c$ and $\delta^{\mathrm{U}}_{\mathrm{G}, m}= 2\pi f_{\mathrm{U}} a_{\mathrm{R,h}} \cos(\psi_{\mathrm{G},m}^{\mathrm{U}}) \sin(\phi_{\mathrm{G},m}^{\mathrm{U}}) / c$ are the spatial frequencies of the AoDs at the RIS for the azimuth and elevation AoDs $\phi_{\mathrm{G},m}^{\mathrm{U}}$ and $\psi_{\mathrm{G},m}^{\mathrm{U}}$, respectively. The RIS-UE channels $\bH_{\mathrm{D}}^{\mathrm{H}}$ and $\bH_{\mathrm{U}}$ can be defined similarly to (\ref{G_D}) and (\ref{G_U}), with $M_{\mathrm{H}}^{\mathrm{D}}$ and $M_{\mathrm{H}}^{\mathrm{U}}$ representing the number of downlink and uplink paths for the RIS-UE link and $a_{\mathrm{U}}$ representing the antenna spacing at the UE.
For the considered scenario we set $M_{\mathrm{G}}^{\mathrm{D}}=M_{\mathrm{G}}^{\mathrm{U}}=M_{\mathrm{H}}^{\mathrm{D}}=M_{\mathrm{H}}^{\mathrm{U}}=5$ and $a_{\mathrm{B}}=a_{\mathrm{R,v}}=a_{\mathrm{R,h}}=a_{\mathrm{U}} =\frac{c}{2 f_{\mathrm{U}}}$.
For the BS-RIS channels, the angles $\zeta_{\mathrm{G},m}^{\mathrm{D}}$ and $\zeta_{\mathrm{G},m}^{\mathrm{U}}$ are randomly generated and uniformly distributed in $[-\pi, \pi)$, and $\phi_{\mathrm{G}, m}^{\mathrm{D}}$, $\psi_{\mathrm{G}, m}^{\mathrm{D}}$, $\phi_{\mathrm{G}, m}^{\mathrm{U}}$, and $\psi_{\mathrm{G}, m}^{\mathrm{U}}$ in $[-\pi/2, \pi/2]$. The angles related to the RIS-UE channels are generated in the same way.
Based on the number of paths, the number of data streams is set to be $N_{\mathrm{s}}^{\mathrm{D}}=N_{\mathrm{s}}^{\mathrm{U}}=5$.

\subsection{Algorithm convergence behavior} \label{sec5_2}
\begin{figure}
	\centering
	\includegraphics[width=1.02\columnwidth]{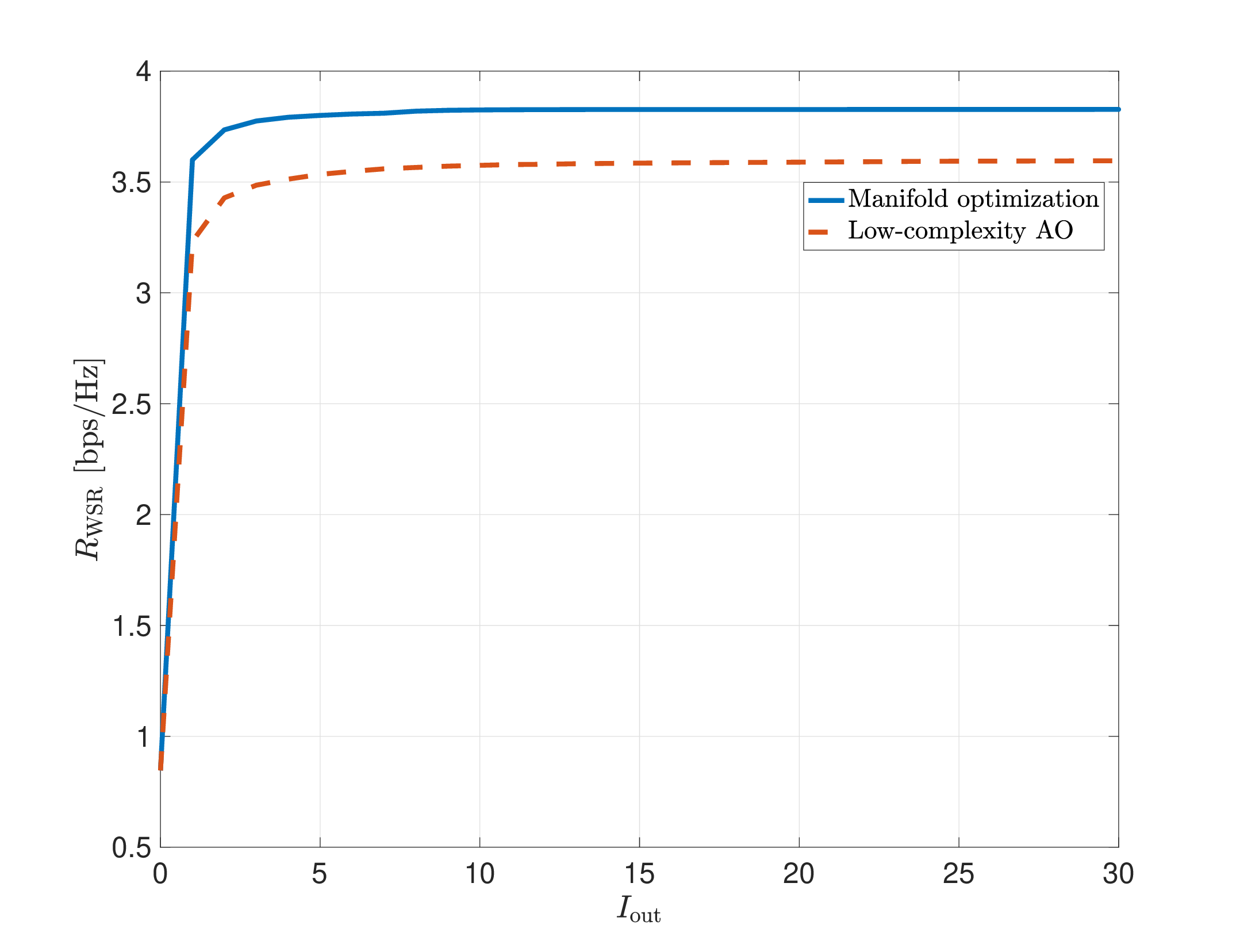}
	\caption{The convergence behavior of proposed algorithms according to the outer iterations.}	\label{Convergence}
\end{figure}
Fig. \ref{Convergence} illustrates the averaged weighted sum-rates of the two proposed algorithms based on 100 independent channel realizations versus the number of outer iterations $I_{\mathrm{out}}$.
Based on the value of $I_{\mathrm{out}}$, the specific number of iterations for both algorithms $I_{\mathrm{out,1}}$ and $I_{\mathrm{out,2}}$ can be determined according to the predefined threshold~$\epsilon$.
It is observed that both algorithms converge to local optimal points. Furthermore, the weighted sum-rates converge within 10 outer iterations, which demonstrates the efficiency of the proposed algorithms in rapidly finding the local optima.
Note that the manifold optimization-based algorithm jointly optimizes all of the RIS reflection coefficients, resulting in a higher weighted sum-rate than the lower-complexity AO algorithm\footnote{In Fig. \ref{Convergence}, we only examine the convergence behavior for the outer iterations. In the manifold-based optimization algorithm, the reflection coefficients are optimized numerically, and the convergence behavior for the inner iterations is guaranteed \cite{Book:manifold}. In the low-complexity AO approach, each reflection coefficient is updated using a closed-form expression, and thus the inner iterations are not needed.}.

\subsection{Performance comparison} \label{sec5_3}
Here we compare the performance of the proposed algorithms against the following baseline approaches:
\begin{itemize}
	\item One-way AO \cite{Zhang:2020}: This scheme can be viewed as a special case of the low-complexity AO approach, where the reflection coefficients at the RIS are optimized to only maximize either the downlink or uplink rate. In one-way AO (DL), the reflection coefficients are optimized solely for maximizing the downlink system performance, while ignoring the uplink. Conversely, for one-way AO (UL), the reflection coefficients are optimized only for the uplink. The remainder of the algorithm is the same as for the low-complexity AO approach, i.e., the precoding matrices $\bF_{\mathrm{D}}$ and $\bF_{\mathrm{U}}$ are computed based on the corresponding updated reflection~coefficients. Note that we simultaneously consider the downlink and uplink performance as long as the actual weight coefficient is in the range $\eta \in (0,1)$.
    \item AO with separated elements: In this approach, the $L$ RIS elements are partitioned into two disjoint sets of size $L/2$. The reflection coefficients in one set are optimized using one-way AO (DL) to support only the downlink, while the other set is optimized using one-way AO (UL) for only the uplink.
	\item Truncated-SVD-based-beamforming (T-SVD-BF) \cite{Wang:2021}: This scheme approximates the singular values of the effective channels $\bH_{\mathrm{eff,D}}$ and $\bH_{\mathrm{eff,U}}$ in terms of the reflection coefficients at the RIS.
	The manifold optimization is applied to update the reflection coefficients, and the precoding matrices are subsequently optimized without any alternating process.
	Note that although this scheme originally targets the downlink design,
	it is also applicable for the joint downlink and uplink case since the computation of the Euclidean gradient for the objective function with both the downlink and uplink can be extended in a straightforward way for the manifold optimization.
	\item Random phase shifts: The phase shifts at the RIS are randomly and uniformly generated in $[0, 2\pi)$, and the precoding matrices are optimized based on eigenmode transmission.
\end{itemize}

\begin{figure}
	\centering
	\includegraphics[width=1.02\columnwidth]{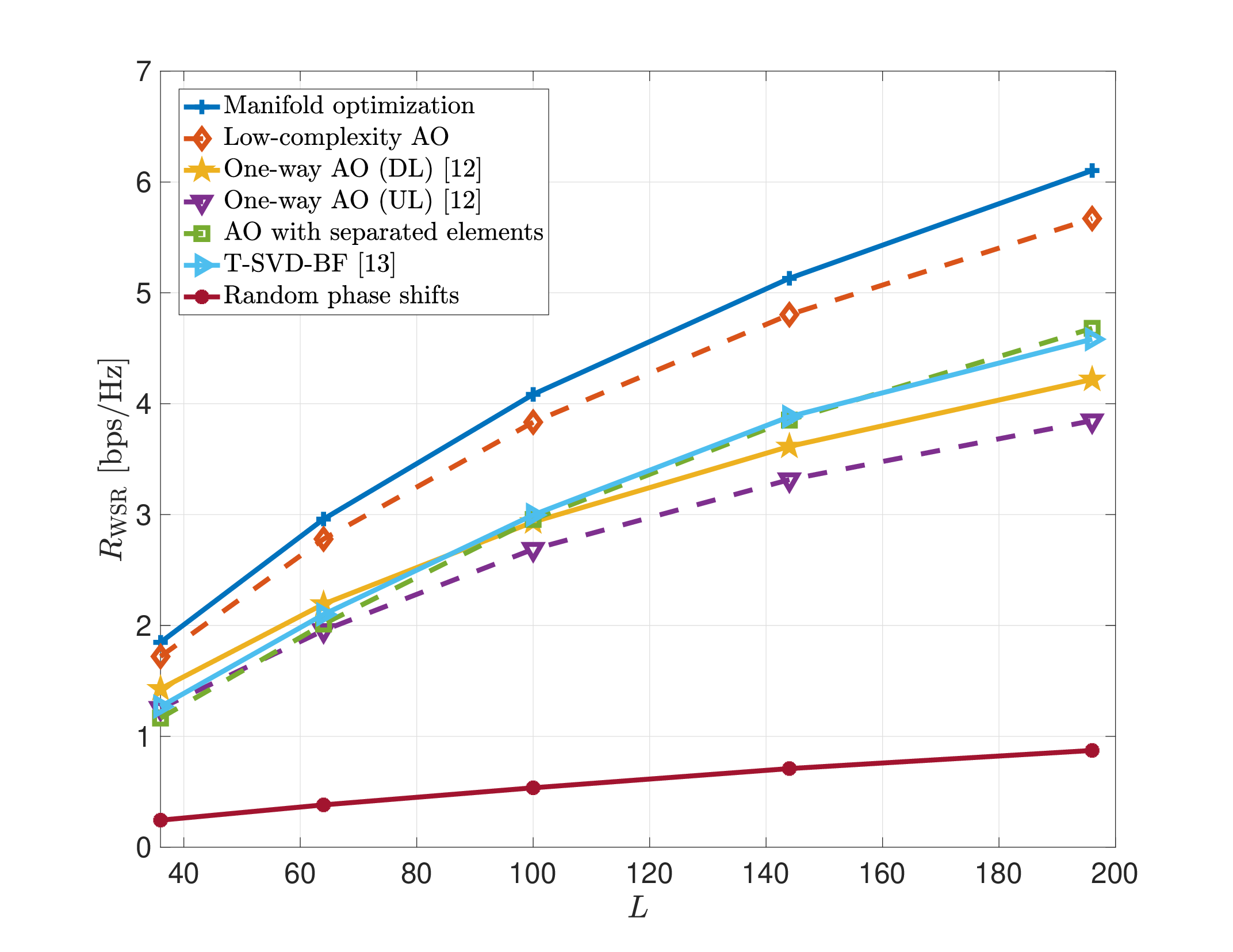}
	\caption{Weighted sum-rate versus the number of RIS elements.} \label{WSR_RIS_elements}
\end{figure}

Fig. \ref{WSR_RIS_elements} shows the weighted sum-rate versus the number of RIS elements $L$ with $L_{\mathrm{h}}=L_{\mathrm{v}}$.
It is observed that the two proposed algorithms achieve the highest weighted sum-rates regardless of the value of $L$, which emphasizes the significance of jointly maximizing the downlink and uplink system performance.
When $L$ is small, the performance gap between the proposed algorithms becomes relatively small due to the resulting reduction in the channel gains, and the approximation used in the low-complexity AO technique works well in this case. Also for small $L$, although T-SVD-BF jointly considers the downlink and uplink when optimizing the RIS reflection coefficients, the performance degradation incurred by the approximation of the singular values of the effective channels leads to a smaller weighted sum-rate than optimizing the downlink only.
Nevertheless, as $L$ increases, T-SVD-BF outperforms the one-way AO-based algorithm since the approximation becomes more accurate, making the joint optimization more effective.
While the AO with separated elements approach shows a higher weighted sum-rate than the one-way AO-based algorithms for large $L$, the set of reflection coefficients optimized for only one direction may cause severe interference in the opposite direction, leading to a performance degradation compared to the proposed algorithms.

\begin{figure}
	\centering
	\includegraphics[width=1.02\columnwidth]{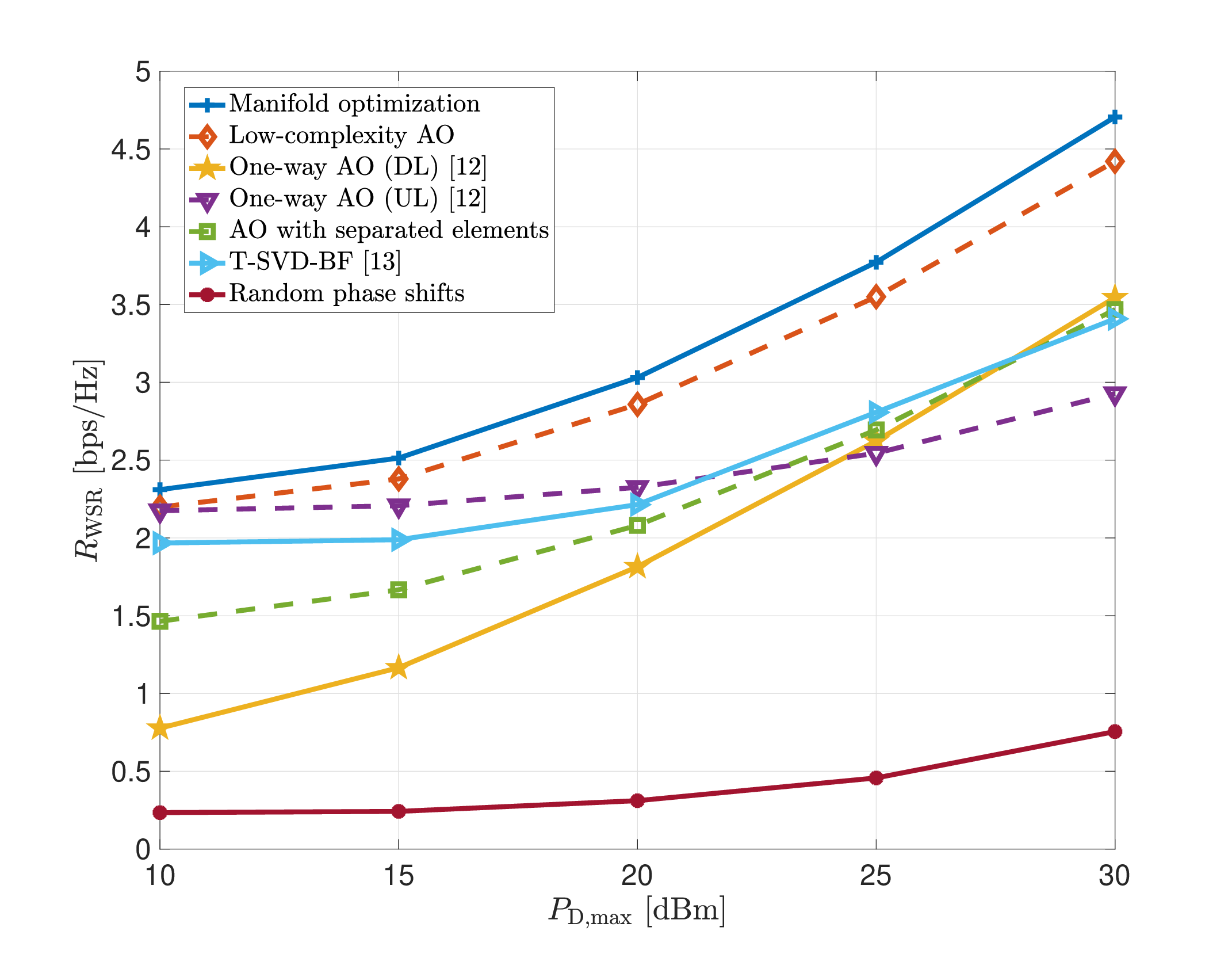}
	\caption{Weighted sum-rate versus the downlink transmit power.} \label{WSR_DL_power}
\end{figure}

Fig. \ref{WSR_DL_power} compares the weighted sum-rate versus $P_{\mathrm{D,max}}$ for fixed $P_{\mathrm{U,max}}=\mbox{23 dBm}$. The proposed algorithms again show the highest weighted sum-rates regardless of the value of $P_{\mathrm{D,max}}$, which illustrates their versatility. 
For small $P_{\mathrm{D,max}}$, the performance gap between the proposed algorithms and the one-way AO (UL)-based algorithm is small since it is better to focus on the uplink only when the downlink power is small.
Similarly, it is observed that the performance of the one-way AO (DL)-based algorithm, which only focuses on the downlink, improves with increasing $P_{\mathrm{D,max}}$.
Although the AO with separated elements approach outperforms the one-way AO-based algorithms under comparable $P_{\mathrm{D,max}}$ and $P_{\mathrm{U,max}}$, the performance gap between the proposed algorithms remains approximately constant, which emphasizes the importance of simultaneously considering both the downlink and uplink in optimizing the reflection coefficients.
 
\subsection{Rate region comparison} \label{sec5_4}

\begin{figure}
	\centering
	\includegraphics[width=1.02\columnwidth]{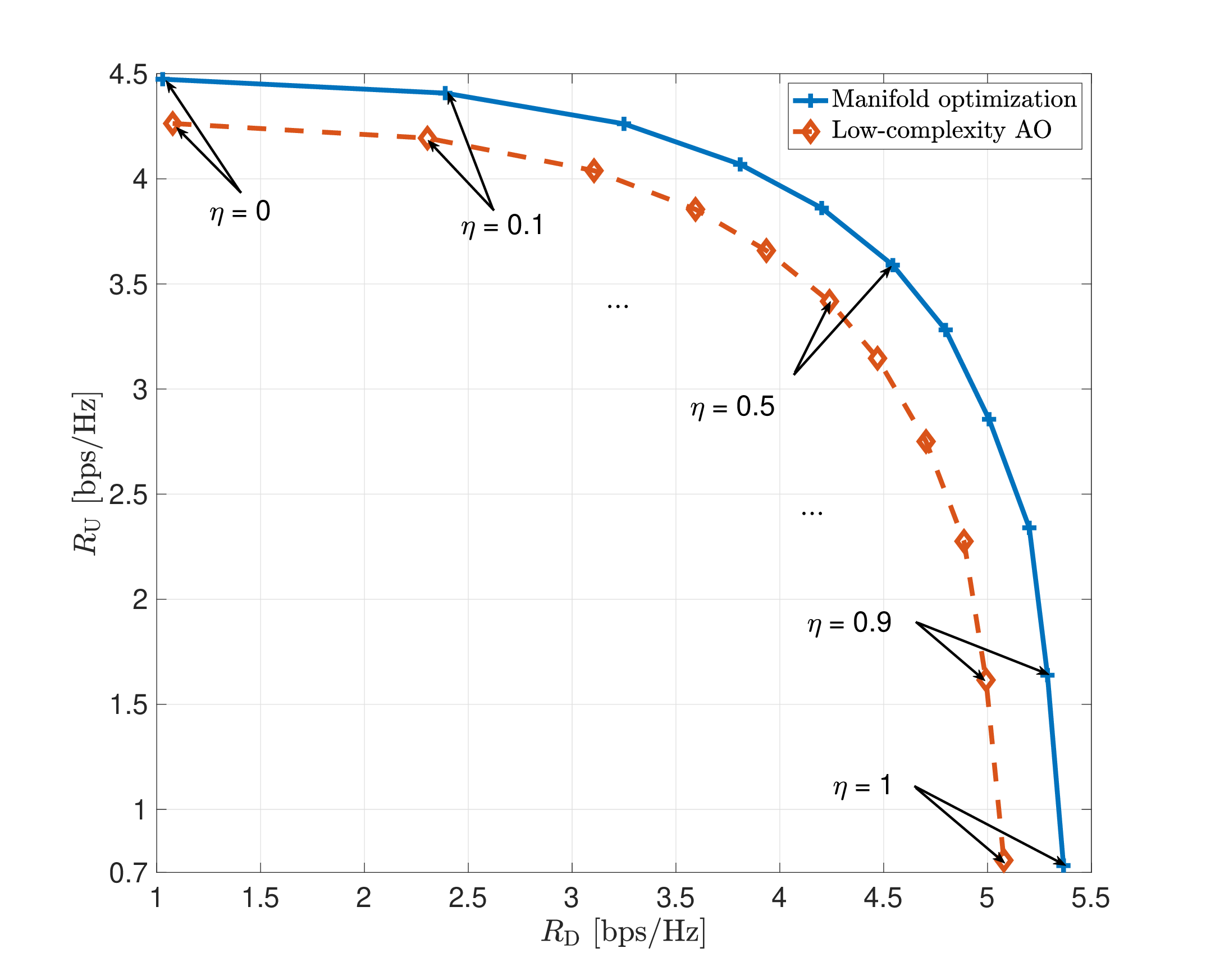}
	\caption{The downlink and uplink rate regions for the proposed algorithms.} \label{Rate_region}
\end{figure}

Lastly we investigate the rate regions for the proposed algorithms obtained by varying $\eta \in [0,1]$. The results are shown in Fig. \ref{Rate_region}, where we see that the manifold optimization-based algorithm achieves higher rates in all cases than the lower-complexity AO algorithm, which is consistent with the previous results. The maximum sum-rate $R_{\mathrm{D}}+R_{\mathrm{U}}$ is achieved when $\eta$ is around 0.5 due to the comparable channel gains and transmit powers for the downlink and uplink. The figure clearly shows that the proposed algorithms can efficiently balance the downlink and uplink performance by adjusting~$\eta$.

\section{Conclusion} \label{conclusion}
In this paper, we investigated RIS-aided FDD SU-MIMO systems, and formulated a joint optimization framework to maximize a weighted sum-rate for the downlink and uplink transmissions.
By adopting an AO algorithm, the precoding matrices at the BS and UE are updated through eigenmode transmissions.
To optimize the reflection coefficients at the RIS, techniques based on manifold optimization and a lower-complexity AO are developed.
Numerical results demonstrated that the proposed algorithms converge quickly and achieve better performance than existing benchmark schemes.
Possible future research directions include extending the joint optimization framework to multi-user systems and developing optimization techniques that consider imperfect CSI.

\appendix{
	\subsection{Proof of Proposition 1}
	To derive (\ref{grad_R_D_H_eff_D}), we use the following theorem from \cite{Palomar:2006}.
	
	\begin{theorem} \label{Appendix_theorem}
		For an arbitrary deterministic matrix $\bH$, an arbitrarily distributed input $\bx$, and Gaussian noise $\bn$ that is independent of $\bx$ with a normalized noise covariance matrix (e.g., an identity matrix), the following equation is~satisfied:
		\begin{equation}
			\nabla_{\bH} I(\bx ; \bH\bx + \bn) = \bH \bE,
		\end{equation}
		where $I(\bx;\by)= \log\det\left(\bI+\bH\bSigma_{\bx}\bH^{\mathrm{H}}\right) $ is the mutual information between the transmit signal $\bx$ and the received signal $\by$, and $\bE = (\bSigma_{\bx}^{-1} + \bH^{\mathrm{H}}\bH)^{-1}$ is the minimum mean squared error (MMSE) matrix with input covariance matrix $\bSigma_{\bx}$.
	\end{theorem}
	
	From (\ref{y_D}), the downlink received signal can be represented by $\by_{\mathrm{D}} = \bH_{\mathrm{eff,D}} \bF_{\mathrm{D}} \bs_{\mathrm{D}} + \bn_{\mathrm{D}} = \tilde{\bH}_{\mathrm{eff,D}} \bs_{\mathrm{D}} + \bn_{\mathrm{D}}$. Following Theorem \ref{Appendix_theorem}, the Euclidean gradient of $R_{\mathrm{D}}$ with respect to $\tilde{\bH}_{\mathrm{eff,D}}$ is given by
	\begin{equation} \label{Appendix_A_2}
		\nabla_{\tilde{\bH}_{\mathrm{eff,D}}} R_{\mathrm{D}} = \frac{1}{\ln2\cdot \sigma_{\mathrm{D}}^2} \tilde{\bH}_{\mathrm{eff,D}} \bE,
	\end{equation}
	where $\bE = \left( \bI_{N_{\mathrm{s}}^{\mathrm{D}}} + \frac{1}{\sigma_{\mathrm{D}}^2 } \tilde{\bH}_{\mathrm{eff,D}}^{\mathrm{H}} \tilde{\bH}_{\mathrm{eff,D}}\right)^{-1}$ is the MMSE matrix. To obtain $\nabla_{\bH_{\mathrm{eff,D}}} R_{\mathrm{D}}$ from (\ref{Appendix_A_2}), 
	we use the following lemma from \cite{Palomar:2006}.
	
	\begin{lemma} \label{Appendix_lemma}
		Let $f$ be a scalar real-valued function, which depends on $\bB$ through $\bH=\bA \bB \bC$, where $\bA, \bC$ are arbitrary fixed matrices. Then, the following equation holds:
		\begin{equation}
			\nabla_{\bB}f = \bA^{\mathrm{H}} \cdot \nabla_{\bH}f \cdot \bC^{\mathrm{H}}.
		\end{equation}
	\end{lemma}
    Plugging in
	$\bH = \tilde{\bH}_{\mathrm{eff,D}}, \bA=\bI_{K}, \bB = \bH_{\mathrm{eff,D}}$, and $\bC = \bF_{\mathrm{D}}$ to Lemma \ref{Appendix_lemma}, $\nabla_{\bH_{\mathrm{eff,D}}} R_{\mathrm{D}}$ is given by
	\begin{align}
		\nabla_{\bH_{\mathrm{eff,D}}}R_{\mathrm{D}} &= \nabla_{\tilde{\bH}_{\mathrm{eff,D}}}R_{\mathrm{D}} \cdot \bF_{\mathrm{D}}^{\mathrm{H}} \\&= \frac{1}{\ln2\cdot \sigma_{\mathrm{D}}^2} \tilde{\bH}_{\mathrm{eff,D}} \bE \bF_{\mathrm{D}}^{\mathrm{H}},
	\end{align}
	which finishes the proof.
	
	\subsection{Derivation of (\ref{f_prime_ell})}
	From (\ref{f_ell}) and (\ref{EVD_AB}), $f^{\prime}_{\mathrm{D},\ell}(\theta_{\ell})$ can be rewritten as
	\begin{align} \label{Appendix_B}
		&f^{\prime}_{\mathrm{D},\ell}(\theta_{\ell}) \nonumber
		\\&=\log_2 \det(\bI_{K} + \theta_{\ell} \bA_{\mathrm{D},\ell}^{-1} \bB_{\mathrm{D},\ell}+\theta_{\ell}^* \bA_{\mathrm{D},\ell}^{-1} \bB_{\mathrm{D},\ell}^{\mathrm{H}})  \nonumber
		\\& = \log_2 \det ( \bI_K + \theta_{\ell} \bU_{\mathrm{D},\ell} \bLambda_{\mathrm{D},\ell}  \bU_{\mathrm{D},\ell}^{-1} \nonumber
		\\& \quad \enspace + \theta_{\ell}^* \bA_{\mathrm{D},\ell}^{-1} (\bU_{\mathrm{D},\ell}^{-1})^{\mathrm{H}} \bLambda_{\mathrm{D},\ell}^{\mathrm{H}} \bU_{\mathrm{D},\ell}^{\mathrm{H}} \bA_{\mathrm{D},\ell} ) \nonumber
		\\& \mathop = \limits^{(a)} \log_2 (\det(\bU_{\mathrm{D},\ell}^{-1})  \det(\bI_K + \theta_{\ell} \bU_{\mathrm{D},\ell} \bLambda_{\mathrm{D},\ell}  \bU_{\mathrm{D},\ell}^{-1} \nonumber
		\\& \quad \enspace + \theta_{\ell}^* \bA_{\mathrm{D},\ell}^{-1} (\bU_{\mathrm{D},\ell}^{-1})^{\mathrm{H}} \bLambda_{\mathrm{D},\ell}^{\mathrm{H}} \bU_{\mathrm{D},\ell}^{\mathrm{H}} \bA_{\mathrm{D},\ell}) \det(\bU_{\mathrm{D},\ell}) ) \nonumber
		\\& = \log_2 \det( \bI_K+\theta_{\ell} \bLambda_{\mathrm{D},\ell}+\theta_{\ell}^* \bC_{\mathrm{D},\ell}^{-1} \bLambda_{\mathrm{D},\ell}^{\mathrm{H}} \bC_{\mathrm{D},\ell} ) \nonumber
		\\& = \log_2 \det(\bI_K+\theta_{\ell}\bLambda_{\mathrm{D},\ell}+\theta_{\ell}^* \lambda_{\mathrm{D},\ell}^* \bc_{\mathrm{D},\ell} (\bc_{\mathrm{D},\ell}^{\prime})^{\mathrm{T}} ) \nonumber
		\\& \mathop = \limits^{(b)} \log_2 \det(1+ \theta_{\ell}^* \lambda_{\mathrm{D},\ell}^* (\bc_{\mathrm{D},\ell}^{\prime})^{\mathrm{T}} (\bI_K+\theta_{\ell} \bLambda_{\mathrm{D},\ell})^{-1} \bc_{\mathrm{D},\ell} ) \nonumber
		\\& \quad \enspace + \log_2 \det(\bI_K + \theta_{\ell} \bLambda_{\mathrm{D},\ell}) \nonumber
		\\& \mathop = \limits^{(c)} \log_2 \left( \left(1+\theta_{\ell}^* \lambda_{\mathrm{D},\ell}^*-\frac{ c^{\prime}_{\mathrm{D},\ell1} c_{\mathrm{D},\ell1} \vert \lambda_{\mathrm{D},\ell} \vert^2 }{1+\theta_{\ell} \lambda_{\mathrm{D},\ell}} \right)(1+\theta_{\ell} \lambda_{\mathrm{D},\ell}) \right) \nonumber
		\\& = \log_2(1+\vert \lambda_{\mathrm{D},\ell} \vert^2 (1-c^{\prime}_{\mathrm{D},\ell1} c_{\mathrm{D},\ell1})+2\mathrm{Re}(\theta_{\ell} \lambda_{\mathrm{D},\ell})),
	\end{align}
	where (a) holds since $\det(\bU_{\mathrm{D},\ell}^{-1}) \det(\bU_{\mathrm{D},\ell})=1$, (b) is derived from the fact that $\det(\bA \bB)=\det(\bA) \det(\bB)$ and $\det(\bI_m + \bC \bD) = \det(\bI_n + \bD \bC)$ for $\bC \in \mathbb{C}^{m \times n}$ and $\bD \in \mathbb{C}^{n \times m}$, and (c) holds since $(\bc_{\mathrm{D},\ell}^{\prime})^{\mathrm{T}} \bc_{\mathrm{D},\ell}=1$ and $\vert \theta_{\ell} \vert^2 =1$. Similarly, $f^{\prime}_{\mathrm{U},\ell}(\theta_{\ell})$ in (\ref{f_prime_ell}) can be derived by following the procedure in~(\ref{Appendix_B}).
}

\bibliographystyle{IEEEtran}
\bibliography{refs_all}

\end{document}